\documentclass[11pt,a4paper]{article}
\pdfoutput=1

\usepackage{bm}
\usepackage{amsmath,amssymb}
\usepackage{cite}

\usepackage{graphicx}
\usepackage{color}
\usepackage{upgreek}
\usepackage[force]{feynmp-auto}
\usepackage{enumitem}

\usepackage{hyperref} 
\hypersetup{colorlinks=true, citecolor=blue, filecolor=black, linkcolor=blue, urlcolor=blue, pdfpagemode=UseNone}

\numberwithin{figure}{section}
\numberwithin{equation}{section}

\makeatletter
\def\namedlabel#1#2{\begingroup
    \textbf{#2.}%
    \def\@currentlabel{\textbf{#2}}%
    \phantomsection\label{#1}\endgroup
}
\makeatother

\newcommand{\nocontentsline}[3]{}
\newcommand{\toclesslab}[3]{\bgroup\let\addcontentsline=\nocontentsline#1{#2\label{#3}}\egroup}

\newcommand{\be}{\begin{equation}}
\newcommand{\ee}{\end{equation}}
\newcommand{\bea}{\begin{eqnarray}}
\newcommand{\eea}{\end{eqnarray}}
\def\beal#1\eeal{\begin{align}#1\end{align}}   
\def\besp#1\eesp{\begin{multline}#1\end{multline}}

\usepackage[normalem]{ulem}

\newcommand{\prop}{\triangle}
\newcommand{\propH}{\bar{\prop}}
\newcommand{\Ph}{\phi}
\newcommand{\ph}{\Phi}
\newcommand{\morris}{Morris:2018axr}

\newcommand{\cu}[1]{\!#1\!}
\newcommand{\Op}{\mathcal{O}}
\newcommand{\hs}{\hat{s}}
\newcommand{\q}{\text{q}}

\newcommand{\dell}[1]{\stackrel{\rightarrow}{\partial}_{#1}\!}
\newcommand{\delr}[1]{\stackrel{\!\leftarrow}{\partial}_{#1}}

\newcommand{\nn}{\nonumber}
\newcommand{\pa}{\partial}

\newcommand\ie{\textit{i.e.}\ }
\newcommand\eg{\textit{e.g.}\ }
\newcommand\cf{\textit{cf.}\ }

\newcommand{\aka}{{a.k.a.}\ }
\newcommand{\etc}{{\it etc.}\ }
\newcommand{\viz}{{\it viz.}\ }
\newcommand{\half}{\tfrac{1}{2}}

\begin{document}

\begin{titlepage}

\begin{center}
{\huge \bf 
BRST in the Exact RG}

\end{center}
\vskip1cm

\begin{center}
{\bf Yuji Igarashi,$^a$ Katsumi Itoh$^a$ and Tim R. Morris$^b$}
\end{center}

\begin{center}
{$^a$ \it Department of Education, Niigata University, Niigata 950-2181, Japan}\\
{$^b$ \it STAG Research Centre \& Department of Physics and Astronomy,\\  University of Southampton,
Highfield, Southampton, SO17 1BJ, U.K.}\\
\vspace*{0.3cm}
{\tt  itoh@ed.niigata-u.ac.jp, igarashi@ed.niigata-u.ac.jp, T.R.Morris@soton.ac.uk}
\end{center}

\abstract{We show, explicitly within perturbation theory, that the
Quantum Master Equation and the Wilsonian renormalization group (RG)
flow equation can be combined {such} that for the continuum effective
{action}, quantum BRST invariance is not broken by the presence of an
effective ultraviolet cutoff $\Lambda$, despite the fact that the
structure demands quantum corrections that na\"\i vely break the gauge
invariance, such as a mass term for a non-Abelian gauge
field. Exploiting the derivative expansion, BRST cohomological methods
fix the solution up to choice of renormalization conditions, {without
inputting the form of the classical, or bare, interactions}. Legendre
transformation results in an equivalent description in terms of
{solving} the modified Slavnov-Taylor identities and the flow of the
{Legendre effective action under an infrared cutoff $\Lambda$ (\ie
effective average action)}. The flow 
generates a canonical transformation that {automatically} solves the
Slavnov-Taylor identities for the wavefunction renormalization
constants. We confirm this
structure in detail at tree level and one loop. Under flow of $\Lambda$,
the standard results are obtained for the beta function, anomalous
dimension, and physical amplitudes, up to choice of renormalization
scheme.}

\end{titlepage}

\tableofcontents

\newpage

\section{Introduction}

Exact renormalization group (RG) equations \cite{Wilson:1973}, such as
those for a Wilsonian effective action, which are exact equations for
its flow with respect to some effective ultraviolet (UV) cutoff
$\Lambda$ \cite{Wilson:1973,Polchinski:1983gv} (see also
\cite{Wegner:1972ih,Weinberg:1976xy,Morris:1993,Latorre:2000qc,Morris:1998kz,Morris:1999px,Morris:2000fs,Arnone:2002cs,Arnone:2002yh,
Arnone:2003pa,Arnone:2005fb,Arnone:2005vd,
Morris:2005tv,Morris:2006in,Morris:2016nda,Falls:2017nnu,Rosten:2018cyr}),
or those for the {effective average action}, which are exact equations
for {the flow of the one-particle irreducible (1PI) effective action,
equivalently the Legendre effective action} with infrared (IR) cutoff
$\Lambda$ \cite{Nicoll1977,Morris:1993,Wetterich:1992} {(see also
\cite{Weinberg:1976xy,Morris:2015oca,Bonini:1992vh,Morris:1995he,Morgan1991}
and \emph{Note Added} in ref. \cite{Morris:1993})}, provide a powerful
conceptual and practical approach to developing both exact and
approximate continuum solutions in quantum field theory (for reviews see
\eg
refs. \cite{Morris:1998,Aoki:2000wm,Bagnuls:2000,Berges:2000ew,Polonyi:2001se,Pawlowski:2005xe,Arnone:2006ie,Kopietz:2010zz,Rosten:2010vm}).

They are formulated using a cutoff function that suppresses the
corresponding region in momentum space of the quantum field. However for
gauge theories, local symmetry transformations, schematically $\phi(x)
\mapsto \Omega(x)\, \phi(x)$, do not respect such a division of the
Fourier transform, $\phi(p)$, into high and low momentum modes. Then one
is faced with either generalising the cutoff and the flow equations such
that they can respect the local invariance of the quantum field
\cite{Morris:1995he,Morris:1998kz,Morris:1999px,Morris:2000fs,Arnone:2002cs,Arnone:2002yh,Arnone:2003pa,Arnone:2005fb,Arnone:2005vd,
Morris:2005tv,Morris:2006in,Rosten:2018cyr,Falls:2017nnu,Morris:2016nda}
(for alternative approaches see
\cite{Branchina:2003ek,Pawlowski:2003sk,Donkin:2012ud,Wetterich:2016ewc,Wetterich:2017aoy})
with however an attendant increase in complexity, or only recovering
gauge invariance in the limit that the cutoff is removed.  In the former
case computations can be carried out while preserving manifest gauge
invariance, however in the latter case one must first fix a gauge. Then
gauge invariance is replaced by BRST invariance
\cite{Becchi:1974xu,Becchi:1974md,Becchi:1975nq,Tyutin:1975qk}, and the
strategy would be to recover BRST invariance in the limit as the cutoff
is removed.

Even if gauge invariance is preserved, BRST invariance is deformed by quantum corrections.
Taking this into account,  BRST invariance is most effectively expressed on the quantum fields 
through the Quantum Master Equation (QME) \cite{Batalin:1981jr,Batalin:1984jr,Batalin:1984ss,Gomis:1994he},
while for the Legendre effective action, it is expressed through the Zinn-Justin equation \cite{ZinnJustin:1974mc,ZinnJustin:1975wb,ZinnJustin:2002ru}, which in fact is most elegantly expressed as the Classical Master Equation (CME).

When combined with the requirement of locality, the CME can be solved by the elegant and powerful methods of BRST cohomology \cite{Fisch:1989rp,Henneaux:1990rx,Barnich:1993vg,Barnich:1994db} (for reviews see \cite{Henneaux:1992ig,Henneaux:1997bm,Gomez:2015bsa}). 
Of course locality is an important physical requirement, but it is also crucial to realise the power of BRST cohomology. Indeed in the space of non-local functionals, this structure disappears: BRST cohomology is trivial \cite{Barnich:1993vg,Barnich:1994db}, and a non-Abelian BRST algebra can be rewritten as an Abelian one \cite{Batalin:1984ss} (see also \cite{Morris:2018axr}). Fortunately, the assumption of locality is justified when analysing the most general possibilities for interacting gauge invariant systems at the classical level and, in sensible frameworks, when analysing only the structure of UV divergences in a perturbative expansion in $\hbar$ (thus \eg leading to the proof of renormalizability of Yang-Mills theories \cite{ZinnJustin:1974mc,ZinnJustin:1975wb,ZinnJustin:2002ru}).

At first sight then, these methods lose their power when analysing the full quantum corrections, which are inherently non-local. Moreover the Batalin-Vilkovisky measure operator $\Delta$, part of the QME, is  ill-defined without further regularisation \cite{Gomis:1994he,Henneaux:1992ig}, and within the framework of exact RG equations, the BRST invariance is anyway broken until the cutoff is removed (as discussed above). 

Actually, as we will see, all these unpleasant features can be avoided in an especially
natural combination of the QME and Exact RG \cite{Morris:2018axr}. In particular they can be combined in such a way  that they are mutually compatible  \cite{Becchi:1996an,Igarashi:1999rm,Igarashi:2000vf,Igarashi:2001mf,Higashi:2007ax,Igarashi:2007fw,Sonoda:2007av,Sonoda:2007dj,Igarashi:2009tj,Frob:2015uqy}. Then the QME can be satisfied simultaneously with the flow equation. By working with off-shell BRST (thus including also auxiliary fields) the quantum BRST invariance is then not lost in the continuum effective action in the presence of the cutoff,
but gets deformed in a calculable way. This is remarkable given that the cutoff then allows a mass term for the gauge field to be generated by quantum corrections \cite{Ellwanger:1994iz}. Such a term is of course forbidden by the classical BRST algebra (and the CME). However it not only becomes allowed but in fact demanded by the now-well-defined (regularised) measure operator $\Delta$ in this framework.

Also locality is not lost, but is relaxed to the requirement of
quasi-locality which is anyway a fundamental requirement of the
Wilsonian RG in the continuum \cite{Morris:1999px,Morris:2000jj},
corresponding to the existence of a sensible Kadanoff blocking
\cite{Kadanoff:1966wm} (with radius of order $1/\Lambda$). Quasilocality
implies that vertices must have a derivative expansion, corresponding to
a Taylor expansion in dimensionless momenta $p^\mu/\Lambda$.  In our
formulation, $\Delta$ is regularised such that it is well-defined
when acting on arbitrary local functionals,
but the rest of the BRST structure remains unmodified to first order \cite{Morris:2018axr}. As we will see, quasi-locality is then sufficient to regain the full potential of BRST cohomological methods.

\emph{Thus we can say that BRST invariance is not broken by the cutoff, but remains powerfully present in this structure. This paper is devoted to exploiting and making manifest this symmetry structure in (the solution of) continuum non-Abelian gauge theory.}

We stress here and throughout the paper that quasi-locality of the effective action at non-vanishing $\Lambda$ is essential. In particular when solving for either such effective action directly in the continuum limit, it is this that replaces the requirement that there exists a local bare action, 
and it is this that ensures a unique solution for the (non-local) effective action at $\Lambda=0$. 
Indeed the physical vertices are obtained from the  Legendre effective action in the $\Lambda\to0$ limit. They are non-local and (evidently) independent of $\Lambda$. Without the constraint that a derivative expansion exists at non-vanishing $\Lambda$ one could add any $\Lambda$-independent {(and BRST invariant)} non-local term to the continuum effective action as a $\Lambda$-integration constant, so that solutions to the flow equations become under-determined. 
On the other hand, the requirement that the effective action has a derivative expansion at non-vanishing $\Lambda$,  replaces the need to insist on a local bare action: indeed it builds one in since by dimensions, higher derivatives appear weighted by the appropriate inverse power of $\Lambda$, and thus the effective action for finitely varying fields (finite momentum) but $\Lambda\to\infty$, tends to a local action (with divergent bare couplings). 

The problems that arise from the impossibility of imposing BRST invariance on some local bare action  at an initial UV scale $\Lambda=\Lambda_0$ (see \eg \cite{Becchi:1996an,Frob:2015uqy}) are avoided by working directly in the continuum limit ($\Lambda_0\to\infty$) with a finite (and thus renormalized) solution for the effective action at scale $\Lambda$.

For $U(1)$ gauge theories, such as QED, it is possible to give a closed form expression for the resulting renormalized BRST transformations in the presence of a cutoff, in terms of manifestly finite composite operators \cite{Igarashi:2007fw}. In non-Abelian gauge theory it does not appear possible to solve exactly in closed form for the BRST symmetry in terms of manifestly finite quantities.\footnote{The problem is that the corresponding composite operators \cite{Igarashi:2007fw} now require renormalization.} However, such a closed form can be derived within suitable approximations.  We demonstrate this in perturbation theory to one loop, using Yang-Mills theory with general gauge group, and in a general gauge. As we will see, the structure allows us to solve directly for the continuum effective action in a particularly clean way, guided and underpinned at crucial stages by quasi-locality and BRST cohomology. Notably, one can work in the so-called minimal gauge invariant basis, where the antighost and auxiliary field are absent, and the freedom in perturbations is manifestly  independent of the choice of gauge fixing.

The functional RG equation for the Wilsonian effective action is related to that for the 1PI effective action  (cutoff Legendre effective action) via a Legendre transform identity \cite{Morris:1993},\footnote{This powerful relation was also exploited in refs.\cite{Morris:2005ck,Morris:2015oca,Bridle:2016nsu,Morris:2018mhd}. See also \cite{Ellwanger1994a,Rosten:2010pc,Bonini:1992vh,Keller:1990ej}.}  such that the UV cutoff in the former is related to the IR cutoff of the latter. 
Under this transformation one obtains not only the flow equation for this so-called effective average action, but also the so-called modified Slavnov-Taylor identities (mST) \cite{Ellwanger:1994iz} which take the form of the CME (\aka Zinn-Justin equation) together with quantum corrections generated by the cutoff. Thus we find that here too, one can solve particularly cleanly for the renormalized 1PI effective action, in minimal gauge invariant basis, exploiting BRST cohomology and quasi-locality. 

{We show that for both effective actions the form of the classical interacting part is uniquely determined by classical BRST cohomology. This was shown in refs. \cite{Fisch:1989rp,Henneaux:1990rx,Barnich:1993vg,Barnich:1994db,Henneaux:1992ig,Henneaux:1997bm,Gomez:2015bsa}. Here we recover this result in the presence of an effective cutoff. In this way, we derive the solution for either effective action without  \textit{a priori} input of either a classical action or a bare action.}

It is clear that since the Wilsonian action can be reconstructed from the 1PI effective action, but satisfies the QME, BRST symmetry is still not lost {at the quantum level} but deformed by the cutoff in a calculable way. Notably, we show that RG flow generates a particularly simple canonical transformation in which the fields and the so-called antifields renormalize in opposite directions. We show that these renormalizations {automatically} solve for the standard Slavnov-Taylor relations between the wavefunction renormalization factors. In particular these relations thus remain {\emph{unaltered} at the quantum level} by the presence of the cutoff. As we will confirm, the coupling parametrises the only remaining freedom in the solution of the QME/mST. 

We verify that the natural renormalization conditions on the effective action correspond to a finite renormalization of standard schemes. {We then find} that at one loop, the coupling {evolves} with $\Lambda$ according to the standard asympotically free  $\beta$-function {(despite the generation of terms that na\"ively break the gauge invariance).}

The structure of the paper is as follows. In the following section, sec. \ref{sec:wil}, we show explicitly that the QME and Wilsonian flow equation, which can be written as compatible equations for finite renormalized quantities, can also be solved consistently for the Wilsonian effective action in terms of finite {quantities.}
By considering a perturbative expansion together with the derivative expansion, we show in secs. \ref{sec:compatibility} and \ref{sec:five} that the resulting structure, even at a general abstract level, is already sufficient to recover {in this context} much of the standard lore for renormalization of non-Abelian gauge theories. We then go on to develop in full, the simultaneous perturbative solution of the flow equation and QME for the case of Yang-Mills with general gauge group {and in general gauge}. In sec. \ref{sec:leg}, we equivalently solve for the 1PI effective action. We derive from the Legendre transform relation, both its flow equation \cite{Nicoll1977,Morris:1993,Wetterich:1992} and its mST equation \cite{Ellwanger:1994iz}. Although this latter is more awkward than the QME, 
it is more streamlined to solve at higher orders since we then avoid having to compute tree-diagram corrections at intermediate stages. However, we also see explicitly via the Legendre transform relations, how these are the same as found for the Wilsonian effective action. These solutions have $\Lambda$-integration constants which must themselves have a derivative expansion. {They determine the interactions through solving the classical BRST cohomology, while at the quantum level they play} the r\^ole of wavefunction renormalization. In sec. \ref{sec:general}, we show that the mST or QME then enforces that this scaling parametrises a simple canonical transformation which {automatically} solves the corresponding Slavnov-Taylor identities. In sec. \ref{sec:oneloopsols} we compute these and the natural renormalization of coupling, verifying that this yields the standard result for the $\beta$ function but here as a flow of the IR cutoff $\Lambda$, and that the resulting vertices satisfy the mST. Finally in sec. \ref{sec:conclusions}, we summarise and draw our conclusions. In particular we emphasise {once} again that in terms of the Wilsonian effective action, the BRST invariance is unbroken, despite the appearance of for example an effective $\Lambda^2$ mass term.


\section{The Wilsonian flow equation and the QME}
\label{sec:wil}

 In this section we solve for gauge theory in the presence of a cutoff in terms of a Wilson/Polchinski effective action $S$  \cite{Wilson:1973,Polchinski:1983gv}, by solving the Wilsonian flow equation simultaneously with the Quantum Master Equation (QME) \cite{Batalin:1981jr,Batalin:1984jr,Batalin:1984ss,Gomis:1994he}. The latter will imply all the Slavnov-Taylor identities as well as the Ward-Takahashi identities. We first consider the issue in general and then specialise to the case of Yang-Mills theory in four spacetime dimensions.

\subsection{BRST invariance and the QME in the presence of a cutoff function}
\label{sec:QME} 
 
We use the notation of ref. \cite{Igarashi:2009tj}  except for small changes. In particular we 
drop $\Lambda$ subscripts since everything will be evaluated at $\Lambda$. However we follow the formulation given in ref. \cite{Morris:2018axr}. It is based on ref. \cite{Igarashi:2009tj}, except that we work with off-shell BRST so that we can use its idempotency, $s^2=0$, which is the crucial property behind BRST cohomology, and we
use a change of variables to a new `basis'. The relation of this basis to earlier work is briefly reviewed in app. \ref{app:shifted}.

We write the  free Wilsonian effective action as \cite{Morris:2018axr}
\be 
\label{S0} 
S_0[\phi,\phi^*] = \half \phi^A K^{-1} \prop^{-1}_{AB}\,\phi^B +\phi^*_A K^{-1} R_{\ B}^A\phi^B\,,
\ee
$\prop^{-1}_{AB}$ being the differential operators defining the kinetic terms, which are regularised by a smooth (thus Taylor expandable) UV cutoff function $K(p^2/\Lambda^2)$, satisfying the standard requirements $K(0)=1$ and $K(u)\to0$ sufficiently fast as $u\to\infty$ to ensure that all momentum integrals that we encounter are UV regulated (faster than power fall off is sufficient). In addition, as we explain later, see \eqref{propmunu}, in general gauge we need the first derivative to vanish: 
\be 
\label{Kprime}
K'(0)=0\,,
\ee
in order to ensure that the effective action has a derivative expansion. The propagator $\prop^{AB}$ is the inverse of $\prop^{-1}_{AB}$. In practice we will be interested in IR regulated propagators 
\be 
\label{propIR}
\propH^{AB} = \bar{K}\prop^{AB}\,,
\ee
where  $\bar{K}=1-K$. Indeed from the above relations, we see that $\bar{K}$ works as an IR cutoff, satisfying $\bar{K}(0)=0$, $\bar{K}'(0)=0$ and $\bar{K}(u)\to1$ as $u\to\infty$.

The antifields $\phi^*_A$ are sources for the BRST transformations. 
Using these, the free action \eqref{S0} 
incorporates the free (and unregularised) BRST transformations:
\be 
\label{Q0gen}
Q_0 \,\phi^A = R_{\ B}^A\,\phi^B\,.
\ee
The full effective action is written $S= S_0 + S_I$, where $S_I[\phi,\phi^*]$ contains the interactions. The  BRST transformations in the interacting theory
\be 
\label{BRS}
Q\phi^A = (\phi^A,S) = K \frac{\partial_lS}{\partial\phi^*_A}
\ee
then follow from a regularised version of the Batalin-Vilkovisky antibracket \cite{Batalin:1981jr,Batalin:1984jr,Batalin:1984ss,Gomis:1994he}. This latter and the regularised measure operator are given by \cite{Morris:2018axr}:
\be 
\label{QMEbitsReg}
(X,Y) = \frac{\partial_rX}{\partial\phi^A}\,K\frac{\partial_lY}{\partial\phi^*_A}-\frac{\partial_rX}{\partial\phi^*_A}\,K\frac{\partial_lY}{\partial\phi^A}\qquad\mathrm{and}\qquad \Delta X = (-)^{A+1} \frac{\partial_r}{\partial\phi^A}\,K\frac{\partial_r}{\partial\phi^*_A}X\,.
\ee
Here $X$ and $Y$ are arbitrary bosonic or fermionic functionals, and $(-)^A=(-)^{\epsilon_A}$, where $\epsilon_A=1(0)$ if $\phi^A$ is fermionic (bosonic), $\phi^*_A$ having the opposite Grassmann grading. It is to be understood that $K$ carries the momentum of the DeWitt-contracted (anti)fields in \eqref{QMEbitsReg}.  These constructs lead to many powerful identities   \cite{Batalin:1981jr,Batalin:1984jr,Batalin:1984ss,Gomis:1994he}, which continue to hold in the presence of this regularisation, since the identities follow from symmetry and statistics  \cite{\morris}.
Notice that the general form \eqref{BRS} agrees with the free case \eqref{Q0gen}, since the cutoff factors cancel between \eqref{QMEbitsReg} and \eqref{S0} in this case. Also notice that the definition \eqref{BRS} implies that the BRST transformations are right-acting, as in  refs. \cite{Batalin:1981jr,Batalin:1984jr,Batalin:1984ss,Gomis:1994he}. (In ref. \cite{Morris:2018axr}, left-acting transformations were used, as in refs. \cite{Fisch:1989rp,Henneaux:1990rx,Barnich:1993vg,Barnich:1994db,Boulanger:2000rq}. In appendix \ref{app:leftright} we translate between the two.) We will also need the Kozsul-Tate operator, $Q^-$ \cite{koszul1950type,borel1953cohomologie,tate1957homology,Morris:2018axr}. It acts (from the right) on antifields \cite{Fisch:1989rp,Henneaux:1990rx,Barnich:1993vg,Barnich:1994db,Henneaux:1997bm,Gomez:2015bsa,Henneaux:1992ig}:
\be 
\label{KT}
Q^- \phi^*_A = (\phi^*_A,  S) = -K \frac{\partial_lS}{\partial\phi^A}\qquad\implies\qquad Q^-_0\phi^*_A = -\prop^{-1}_{AB}\,\phi^B + (-)^{B}\phi^*_B R^B_{\ A}\,.
\ee

The QME is the statement that the Quantum Master Functional (QMF) vanishes, where the QMF is given by \cite{Batalin:1981jr}
\be 
\label{QMF}
\Sigma = \half (S,S) -\Delta S\,,
\ee
regularised as in \eqref{QMEbitsReg}. 
Of particular importance are perturbations, $S+\varepsilon\,\Op$, that preserve the QME  (where $\varepsilon$ is infinitesimal, and $\Op$ a quasi-local operator integrated over spacetime). This deforms the BRST algebra, allowing us to explore the space of interacting theories whose gauge invariance is smoothly connected to that of the original. Substituting the perturbed action into the QME we have that the operator 
must be (quantum) BRST invariant:
\be 
\label{anihilate}
\hat{s}\,\Op=0\,, 
\ee
\aka closed under $\hat{s}$, where the \emph{full} (or total) quantum BRST transformation,\footnote{{This was called $s$ in \cite{\morris}, which here we reserve for $s=Q+Q^-$.}}
\be 
\label{fullBRS}
\hat{s}\,\Op = (\Op,S) -\Delta\Op  = (Q+Q^-\!-\Delta)\,\Op\,,
\ee
can be shown to be {nilpotent}: $\hat{s}^2=0$ \cite{Batalin:1981jr,Batalin:1984jr,Batalin:1984ss,Gomis:1994he}.
An infinitesimal canonical transformation: 
\be 
\label{sK}
\Op = \hat{s}\, \mathcal{K} = (\mathcal{K},S)-\Delta \mathcal{K}\,,
\ee
is $\hat{s}$-exact, and thus trivially a solution to \eqref{anihilate} and a symmetry of the QME. However such an operator just corresponds to infinitesimal field and source redefinitions: 
\be 
\delta \phi^A = -\varepsilon K\frac{\partial_r \mathcal{K}}{\partial \phi^*_A}\,,\qquad \delta\phi^*_A = \varepsilon K\frac{\partial_r\mathcal{K}}{\partial\phi^A}\,,
\ee
with $-\Delta \mathcal{K}$  corresponding to the Jacobian of the change of variables in the partition function. We are therefore interested only in operators $\Op$ that are closed under $\hat{s}$ but not exact, \ie we want operators that lie in the non-trivial part of the (full) quantum BRST cohomology.

When we develop solutions as an expansion in $\hbar$, and also when we consider the 1PI framework in sec. \ref{sec:leg},  the classical ($\hbar\to0$) equivalents to the above become important. They are the same equations with $S$ replaced by the classical action, $S_\text{cl}$, and with the measure operator switched off. In particular the Classical Master Equation (CME) is the statement that:
\be 
\label{CME}
(S_\text{cl},S_\text{cl}) = 0\,.
\ee
The full classical BRST transformation, $s_\text{cl}$, on an operator $\Op$ is given by the sum of the classical Koszul-Tate operator and classical BRST operator:
\be   
\label{fullBRSC}
s_\text{cl}\,\Op = (Q_\text{cl}+Q^-_\text{cl})\, \Op = (\Op,S_\text{cl})\,.
\ee
Perturbations, $S_\text{cl}+\varepsilon\,\Op$, that preserve the CME are invariant under the full classical BRST transformation: $s_\text{cl}\,\Op = 0$.

\subsection{Compatibility with the Wilsonian flow equation and continuum limit}
\label{sec:compatibility}

As shown in ref. \cite{Morris:2018axr}, in this formulation the flow equation takes the same form as in ref. \cite{Polchinski:1983gv}:
\be 
\label{flow}
\dot{S}_I = -\half\partial^r_A S_I \dot{\propH}^{AB}\partial^l_BS_I +\half (-)^A \dot{\propH}^{AB} \partial^l_B\partial^r_A S_I = \half a_0[S_I,S_I] - a_1[S_I]\,,
\ee
where ($a_0$)$a_1$ is the corresponding (bi)linear functional detailed in the middle equation \cite{Morris:1998kz}. Here $\partial_A \equiv \partial/\partial \phi^A$, and the dot means $\partial_t = -\Lambda \partial_\Lambda$, where $t= \ln(\mu/\Lambda)$, $\mu$ being the usual arbitrary fixed finite energy scale.
With the regularisation \eqref{QMEbitsReg}, the QMF is compatible with the flow equation \eqref{flow} \cite{Morris:2018axr}: 
\be 
\label{compatible}
\dot{\Sigma} = a_0[S_I,\Sigma]-a_1[\Sigma]\,.
\ee
This means that if $\Sigma =0$, then also $\dot{\Sigma}=0$, and thus that if the QME is satisfied at some generic scale $\Lambda$, it remains satisfied on further RG evolution. (In fact \eqref{compatible} coincides with the flow equation for a (composite) operator  \cite{Igarashi:2009tj,Morris:2018axr}.)

On the other hand, we are interested in forming a continuum limit, which
means that in dimensionless variables (and appropriate wavefunction
renormalizations), $S$ tends to a fixed point action in the limit
$\Lambda\to\infty$. Close to the fixed point, $S$ is then given by the
fixed point action plus an expansion to first order over the
(marginally) relevant eigenoperators, with conjugate couplings $g$ which
(in dimensionful terms) are constant and of positive or vanishing
dimension \cite{Wilson:1973,Morris:1998,Morris:2018mhd}. These cases are
not generic points on the flow.  It is not sufficient that the QME is
satisfied at the fixed point, or even at first order, to be able to
conclude from \eqref{compatible} that the QME is satisfied at all points
on the resulting renormalized trajectory. However, developing the
solution as a perturbative expansion over the eigenoperators beyond
first order, we will see that the compatibility of BRST invariance with
the continuum limit is determined only by integration constants
associated with the higher order solutions.\footnote{Interestingly this
leaves open the possibility of extra constraints from non-perturbative
corrections. We will not explore this further here.} This serves to
determine how a cutoff effective action that is invariant under such
non-linear BRST transformations, must have its (marginally) relevant
couplings constrained. 

Let us emphasise that the above procedure can in principle be
carried out around a non-perturbative fixed point (using suitably
modified equations to take into account anomalous dimensions). The key
observation that makes this possible is that close to the fixed point a
perturbative expansion over the (marginally) relevant couplings is
allowed even in this case, since such couplings must be vanishingly
small there. Then the integration constants will be determined in the
way we sketched above.

To carry out such a program explicitly however requires explicit
solutions for a non-perturbative fixed point which also satisfies the
QME, and explicit solutions for the eigenoperators. No exact
non-perturbative solutions are known. Therefore, here we will content
ourselves with demonstrating the structure within perturbation theory
about the Gaussian fixed point.

Thus, consider a renormalized trajectory leaving the Gaussian fixed
point at $\Lambda=\infty$. The Gaussian fixed point corresponds to the
solution $S_I=0$ of \eqref{flow}. The flow is parametrized by the finite
renormalized (marginally) relevant couplings. For simplicity we assume
there is only one coupling $g$.
Since we are studying the flow close to the fixed point, we can expand:
\be 
\label{Sexpand}
S = S_0 + gS_1 +g^2S_2+\cdots\,.
\ee
Here $S_1$ is the eigenoperator conjugate to $g$. When we consider quantum corrections we will need to define $g$ more precisely, through some renormalization condition. As we recall later, depending on our choice of renormalization condition, $g$ can either be a function of $\Lambda$ or independent of it. Here we treat it as $\Lambda$ independent. 
Perturbative expansion of \eqref{flow} then gives 
\be 
\label{flowSn}
\dot{S}_n = \frac12 \sum_{m=1}^{n-1} a_0[S_{n-m},S_m]\,-a_1[S_n]\,.
\ee
Similarly one can expand the QMF as 
\be 
\Sigma = \Sigma_0 + g\Sigma_1 +g^2\Sigma_2+\cdots\,.
\ee
Now $\Sigma_0=0$ follows from the fact that \eqref{Q0gen} is an invariance of the kinetic term in \eqref{S0}.
Then one has from \eqref{compatible} that
\be 
\label{flowSigman}
\dot{\Sigma}_n  =  \sum_{m=1}^{n-1} a_0[S_{n-m},\Sigma_m] -a_1[\Sigma_n]\,.
\ee
From \eqref{QMF} we have that
\be 
\label{Sigman}
\Sigma_n = \hat{s}_0\, S_n\, + \frac12 \sum_{m=1}^{n-1} (S_{n-m},S_m)\,,
\ee
where from \eqref{fullBRS}, $\hat{s}_0=Q_0+Q^-_0-\Delta$ is the free full quantum BRST charge.

Now, from \eqref{flowSn}, the eigenoperator equation is in this notation:
\be 
\label{flowSone}
\dot{S}_1 = -a_1[S_1]\,.
\ee
From \eqref{flowSigman} we see that $\Sigma_1$ also satisfies this equation. From \eqref{Sigman}, $\Sigma_1 = \hat{s}_0\,S_1$. Therefore $\hat{s}_0\,S_1$ must also be an expansion over eigenoperators with constant coefficients. Since we require $\Sigma_1=0$ for the QME to be satisfied, we thus see that the quantum BRST cohomology is to be defined within the space spanned by the eigenoperators with constant coefficients \cite{Morris:2018mhd}.

If we have already determined that the QME is solved up to $\Sigma_m=0$ for $m<n$, then \eqref{flowSigman} reduces to 
\be 
\label{SigmanFreedom}
\dot{\Sigma}_n = -a_1[\Sigma_n]\,.
\ee
Since this is again the eigenoperator equation, it means that the QME has only the possibility to be violated by a linear combination of eigenoperators with constant coefficients. By 
\eqref{Sigman}, this is to be repaired (if possible) adding to $S_n$ a linear combination of eigenoperators with constant coefficients. We thus see that the only freedom to make the QME compatible with a continuum limit solution of the flow equation is that contained within the perturbative development of the BRST cohomology.\footnote{In particular note that from \eqref{flowSn} and \eqref{flowSigman}, we are always free to add to $S_n$ an $\hat{s}_0$-closed linear combination of eigenoperators with constant coefficients. It is this freedom that needs to be fixed by renormalization conditions.}

{Let us stress that this is a very powerful conclusion. It is clear
that in this way we are already recovering the crucial steps in the
proof of renormalizability of gauge theories
\cite{ZinnJustin:1974mc,ZinnJustin:1975wb,ZinnJustin:2002ru}. We will
develop this in the remainder of the paper. However we emphasise that
this structure is here seen to arise in general, for the continuum (and
thus we mean already renormalized) effective action. And note especially
that this structure is seen to operate even with the momentum cutoff
kept in place.}

{Notice that although the equations above have been derived by
expanding perturbatively in $g$, they are still exact in $\hbar$, \ie
non-perturbative in the loop expansion. For example consider the
eigenoperator equation \eqref{flowSone}. On the RHS, $a_1[S_1]$
attaches a loop to the $S_1$ vertex. If the result is non-vanishing,
solutions $S_1$ will be functions of $\hbar$. This dependence may be
Taylor expandable in $\hbar$, or may also be inherently non-perturbative
in $\hbar$. The latter type of eigenoperator is the crucial starting
point in a proposal for a genuinely renormalizable quantisation of
gravity
\cite{Morris:2018mhd,Kellett:2018loq,Morris:2018upm,\morris}. The other
equations, \eqref{flowSn} -- \eqref{SigmanFreedom}, are similarly exact
in $\hbar$. As we have seen they are solved by utilising
$\hat{s}_0$-cohomology, where $\hat{s}_0$ has a linear dependence on
$\hbar$, given exactly by the one-loop measure operator, $\Delta$.}

{Starting in the next section however, we will specialise to
Yang-Mills theory, whose solution may be computed perturbatively in the
loop expansion. In fact in this case the $S_1$ we need to use is purely
classical (\ie carries no $\hbar$ dependence) since $a_1[S_1]$ turns out
to vanish.}

\subsection{Five consequences for Yang-Mills theory}
\label{sec:five}

Let us illustrate {the power of the structure above} by now specialising to the case of Yang-Mills theory in four spacetime dimensions, and to the loop expansion (\ie expansion in $\hbar$). We will need at this stage only that all (anti)fields have at least dimension one (see table \ref{table:ghostantighost}), the coupling $g$ is dimensionless, $\Sigma$ ($S$) has dimension one (zero) overall, and that $S_1$ contains monomials with (at least) three (anti)fields. 

At the classical level, $S=S_{\text{cl}}$, we drop $a_1$. Then from \eqref{SigmanFreedom} the only way the QME (now CME) can be violated is through terms that are independent of $\Lambda$. Since $S$ ($\Sigma$) must have a derivative expansion and since there are no dimensionful parameters other than $\Lambda$ itself, such terms in $S$ ($\Sigma$) are spacetime integrals of local operators of dimension four (five). We therefore conclude that:

\begin{enumerate}
\item[\namedlabel{point1}{C1}]  from \eqref{flowSone} which is now $\dot{S}_{1,\text{cl}}=0$, and $\Sigma_{1,\text{cl}}=0$ which is
\be \label{classicalCohomology}
s_0 \,S_{1,\text{cl}}= 0 \,, \qquad \text{where}\qquad s_0 := Q_0+Q^-_0 \,,
\ee
$S_{1,\text{cl}}$ is a local operator of dimension four, which moreover must be a non-trivial solution of the classical free BRST cohomology ($s_0$ being the full free classical BRST charge);
\item[\namedlabel{point2}{C2}] $S_{2,\text{cl}}$ being formed of tree-level corrections involving $S_1$, has a derivative expansion to all orders, however $\Sigma_{2,\text{cl}}$ must already vanish for all the $\Lambda$ dependence, leaving at most a dimension five local operator that we need to cancel (if possible) with a $\Lambda$-independent dimension four local operator addition to $S_{2,\text{cl}}$;
\item[\namedlabel{point3}{C3}] since the tree expansion, and similarly the antibracket in \eqref{Sigman}, has the property that $S_{n,\text{cl}}$ ($\Sigma_{n,\text{cl}}$) is made of vertices with at least $n\cu+2$ (anti)fields,  no $\Lambda$-independent local operator is possible in $S_{3,\text{cl}}$. Thus any $\Lambda$-independent (and thus dimension five) part of $\Sigma_{3,\text{cl}}$ would represent an obstruction to the classical solution at $O(g^3)$. For all $n\cu>3$, neither $S_{n,\text{cl}}$ nor $\Sigma_{n,\text{cl}}$ can have a $\Lambda$-independent local term, and therefore the $S_{n>3,\text{cl}}$ must already satisfy the CME.
\end{enumerate}

Now consider the quantum level at some fixed order $\ell$ of the loop expansion. As we will review, the solution of the flow equation generates the expected momentum integrals, however with propagators that contain an IR cutoff. Assuming the lower loop orders  have been solved already, since the right hand side generates an extra loop,  \eqref{SigmanFreedom} again reduces to the statement that the only way the QME can be violated is by terms that are independent of $\Lambda$, which thus in $S_n^{(\ell)}$ ($\Sigma_n^{(\ell)}$) correspond to dimension four (five) local operators. We see that the problem of solving simultaneously the flow equation and QME is therefore confined to the question of whether such $\Lambda$-integration constants can be chosen satisfactorily. We have thus shown that at the quantum level:
\begin{enumerate}
\item[\namedlabel{point4}{C4}] the body of the momentum integrals, corresponding to all the non-trivial $\Lambda$ dependence, must already automatically satisfy the QME.
\end{enumerate}

The $\Lambda$-integration constants play the same r\^ole as counterterms in a more standard treatment. Indeed, since the measure operator in \eqref{Sigman} also generates an extra loop, the $\Lambda$-integration constants are parts of $S_n^{(\ell)}$ that are 
determined only up to $s_0$-closed pieces. The freedom in the solution is thus held in pieces that are either proportional to $S_{1,\text{cl}}$ itself, \ie the solution to \eqref{classicalCohomology}, or to $s_0$-closed bilinear terms -- which are therefore contained in the $\Lambda$-independent ($K\mapsto1$) parts  of \eqref{S0}. The latter thus correspond to the freedom to introduce wavefunction renormalizations $Z_i$ such that the CME remains satisfied. {In other words, the $Z_i$ must in fact parametrise a particularly simple form of canonical transformation. As we will show fully in sec. \ref{sec:general}}:
\begin{enumerate}
\item[\namedlabel{point5}{C5}] the $Z_i$ parametrise a canonical transformation that {automatically} solves the standard Slavnov-Taylor identities for the wavefunction renormalization constants.
\end{enumerate}
This freedom in the solution is  fixed by choosing suitable renormalization conditions.

Although powerful, as we have seen these conclusions are achieved while remaining at an abstract level. In the rest of the paper, we will solve for the effective actions $S$ and $\Gamma$, for Yang-Mills at tree-level and one-loop level up to $O(g^3)$, thus providing the explicit form that confirm the above results, and allowing the structure to be understood in full in a concrete example. In particular we will see explicitly that, despite the apparent explicit breaking of gauge invariance by the cutoff, the QME and its implied BRST invariance remain satisfied. They in fact now demand quantum corrections that na\"\i vely break the gauge invariance, such as a mass term for the gauge field, in a way that ensures a smooth limit to the standard results once the cutoff is removed. They also ensure the standard $\beta$-function, which can even be recovered from a flow of $g(\Lambda)$ with respect to cutoff $\Lambda$ itself.

\subsection{Free action, gauge fixed and (minimal) gauge invariant basis}
\label{sec:S0}

Before introducing gauge fixing, the free action is in the so-called minimal gauge invariant basis \cite{Batalin:1981jr,Batalin:1984jr,Batalin:1984ss,Gomis:1994he}:
\be 
\label{Sfreegimin}
S_0 
= \half \partial_\mu a_\nu K^{-1} \partial_\mu a_\nu -\half \partial\!\cdot\! a K^{-1} \partial\!\cdot\! a + a^*_\mu K^{-1} \partial_\mu c\,.
\ee
Here $a_\mu$ is the Yang-Mills gauge field, $c$ the ghost field,
and $a^*_\mu$ is the source (antifield) for the free BRST transformation on $a_\mu$.
Our notation is such that any expression for an action functional should be understood to appear inside the braces in 
\be \label{notation} 2\, \text{tr} \int_x \{ \cdots \}\,.\ee 
The (anti)fields depend on $x$,  and unless explicitly stated that they are contracted into the generators of the gauge group, \eg $c=c^a T^a$. The factor of $2$ in \eqref{notation} compensates for the fact that the generators are ortho{normaliz}ed to $\text{tr}( T^a T^b) = \delta^{ab}/2$, as usual.

We will see shortly that we can in fact work in this minimal gauge invariant basis, where expressions are simplest, and it is clear the extent to which calculations are independent of gauge fixing. In this basis, the only extra field we will need is $c^*$, the source for BRST transformations of $c$ that appear at the interacting level. 

In order to gauge fix however, we must first add the auxiliary field $b$ that allows off-shell BRST invariance, and $\bar{c}^*$ which sources BRST transformations of the antighost, $\bar{c}$. {This takes us to the so-called extended (or non-minimal) gauge invariant basis. The free action is then} written as:
\be 
\label{Sfreegi}
S_0 |_\text{gi} = \half \partial_\mu a_\nu K^{-1} \partial_\mu a_\nu -\half \partial\!\cdot\! a K^{-1} \partial\!\cdot\! a + a^*_\mu K^{-1} \partial_\mu c + \frac{\xi}{2} b K^{-1} b + \bar{c}^* K^{-1} b\,,
\ee
where $\xi$ will become the gauge fixing parameter. Recall that $S_0 \equiv S_{0,\Lambda}$ so everything here is at a physical scale and renormalized, thus $\xi$ is renormalized and not bare for example.

For the flow equation \eqref{flow} we need propagators, which we get by working in gauge fixed basis. This is implemented by a (finite) quantum canonical transformation \cite{Siegel:1989ip,Siegel:1989nh,VanProeyen:1991mp,Bergshoeff:1991hb,Troost:1994xw,Gomis:1994he} that affects only the antifields:
\be 
\label{canontogf}
\phi^*_A |_\text{gf} = \phi^*_A |_\text{gi} + \partial^r_A \Psi\,.
\ee
Unlike the above references we do not then set the antifields to zero, since we will need them to solve for the BRS transformations  \cite{Morris:2018axr}. We take the gauge fermion to be 
\be 
\label{Psi}
\Psi = - i \bar{c}\, \partial\!\cdot\! a\,.
\ee
Thus explicitly
\beal
\bar{c}^*  |_\text{gf}  &= \bar{c}^*  |_\text{gi} - i \partial\!\cdot\! a\,, \nonumber\\
a^*_\mu |_\text{gf} & = a^*_\mu |_\text{gi} + i \partial_\mu \bar{c}\,.
\label{gfgi}
\eeal
Then the free action in gauge fixed basis is:
\be 
\label{Sfreegf}
S_0 |_\text{gf} = S_0 |_\text{gi}  - i \partial\!\cdot\! a K^{-1} b -i \partial_\mu\bar{c} K^{-1} \partial_\mu c\,.
\ee
Leaving out the regulator, and inverting the $\prop^{-1}_{AB}$ in \eqref{Sfreegf}, gives the propagators $\prop^{AB}$:
\beal
\langle c^a(p)\, \bar{c}^b(-p) \rangle &= i\delta^{ab} \prop(p)\,,\quad\text{where}\quad \prop(p) := 1/{p^2}\,,\nonumber\\
\langle a^a_\mu(p)\, {a}^b_\nu(-p) \rangle &= \delta^{ab} \prop_{\mu\nu}(p)\,, \quad\text{where}\quad \prop_{\mu\nu}(p):=\frac{1}{p^2}\left(\delta_{\mu\nu} + (\xi-1)\frac{p_\mu p_\nu}{p^2}\right)\,,\nonumber\\
\langle b^a(p)\, {a}^b_\mu(-p) \rangle &= - \langle a^a_\mu(p)\, {b}^b(-p) \rangle = \delta^{ab} p_\mu\prop(p)\,,\nonumber\\
\langle b^a(p)\, {b}^b(-p) \rangle &= 0\,.
\label{propagators}
\eeal

By definition, since \eqref{gfgi} is a quantum canonical transformation, the QMF, \viz \eqref{QMF}, is invariant. As we saw in sec. \ref{sec:five}, the lowest order interaction $S_1=S_{1,\text{cl}}$ is given by a $\Lambda$-independent non-trivial solution of the BRST cohomology. We can thus determine $S_{1,\text{cl}}$ in gauge invariant minimal basis, which is manifestly independent of choice of gauge and uses the minimum number of (anti)fields. This means in particular that $S_{1,\text{cl}}$ can be chosen to be independent of $b$ and $\bar{c}^*$.
To find the higher orders in \eqref{Sexpand}, we then need the flow equation. Since the flow equation is not invariant under the transformation \eqref{gfgi} one strategy would be first to map $S_{1,\text{cl}}$ to gauge fixed basis. Once we have the result we need from the flow equation, we can map back to gauge invariant basis, where expressions are simpler, using \eqref{gfgi}. This was the strategy used in ref. \cite{Morris:2018axr}.

However we can streamline this procedure, since we can already deduce the effect of mapping between the different bases. 
First we note that from \eqref{gfgi},  $S_{1,\text{cl}}$ in gauge fixed basis will still not depend on $b$ or $\bar{c}^*$, and depends on $\bar{c}$ only through the combination $a^*_\mu - i \partial_\mu\bar{c}$. By iteration, using the flow equation \eqref{flow}, these properties are inherited by all the higher order interactions. Mapping back to gauge invariant basis using \eqref{gfgi}, we therefore see that $S_I$ will not depend on $b$, $\bar{c}^*$ or $\bar{c}$. This means in particular that the full $S_I$ remains in \emph{minimal} gauge invariant basis. However from \eqref{gfgi}, we must replace field derivatives in \eqref{flow} with
\be 
\label{gfgipartials}
\frac{\partial}{\partial{a}_\mu} \Big|_\text{gf}= \frac{\partial}{\partial{a}_\mu} -i\partial_\mu \frac{\partial}{\partial{\bar{c}}^*}\,,\quad
\frac{\partial}{\partial\bar{c}} \Big|_\text{gf} = \frac{\partial}{\partial{\bar{c}}} + i\partial_\mu\frac{\partial}{\partial{a}^*_\mu}\,,\quad \frac{\partial}{\partial{a}^*_\mu} \Big|_\text{gf}= \frac{\partial}{\partial{a}^*_\mu}\,,\quad \frac{\partial}{\partial\bar{c}^*} \Big|_\text{gf}=\frac{\partial}{\partial{\bar{c}}^*} \,,
\ee 
where the differentials on the RHS are in gauge invariant basis. Since $S_I$ does not depend on either $\bar{c}^*$ or $\bar{c}$, acting on $S_I$ these derivatives are effectively unchanged except for the replacement
\be 
\label{gibasisDeriv}
\frac{\partial}{\partial\bar{c}}  \Big |_\text{gf}\ \equiv\   i\partial_\mu\frac{\partial}{\partial{a}^*_\mu} \Big |_\text{gi}\,.
\ee 
This is equivalent for the purposes of taking a $\bar{c}$ derivative, to temporarily replacing $a^*_\mu$ by $a^*_\mu - i \partial_\mu\bar{c}$. Either way, 
this one replacement ensures that we get the right result from the flow equation when working in minimal gauge invariant basis.

Since it simplifies the results, for our calculations we will therefore work in this basis. Although  $S_{1,\text{cl}}$ is then entirely independent of choice of gauge,
the higher order corrections do depend on it. Indeed this is clear from the fact that the gauge field propagator in \eqref{propagators}, depends on $\xi$. 

As already stated in \eqref{propIR} and already clear to some extent from \eqref{flow}, the propagators that appear in vertices will be IR regulated by $\bar{K}(p^2/\Lambda^2)$. As recalled in the Introduction, the effective actions must have the property that they have a derivative expansion for $\Lambda>0$.  Non-trivial solutions to the effective actions will only  have a derivative expansion if the propagators also have a derivative expansion. This means the propagators must have a Taylor expansion in $p^\mu$. Since for the gauge   fields
\be 
\label{propmunu}
\propH_{\mu\nu}(p) = 
\bar{K}(p^2/\Lambda^2)\, \prop_{\mu\nu}(p) 
= - \frac{K'(0)}{\Lambda^2} \left(\delta_{\mu\nu} + (\xi-1)\frac{p_\mu p_\nu}{p^2}\right) + \hbox{Taylor series}\,,
\ee
we see that this property is violated for the gauge propagator outside Feynman gauge unless we impose condition \eqref{Kprime}, $K'(0)=0$.

\subsection{BRST algebra}
\label{sec:brst}

\begin{table}[ht]
\begin{center}
\begin{tabular}{|c|c|c|c|c|c|c|}
\hline
 & $\epsilon$ & gh \# & ag \# & pure gh \#  & dimension \\
\hline\hline
$a_\mu$ & 0 & 0 & 0 & 0 & $1$ \\
\hline
$c$ & 1 & 1 & 0 & 1 & $1$ \\
\hline \hline
$\bar{c}$ & 1 & -1 & 1 & 0 & $1$  \\
\hline
$b$ & 0 & 0 & 1 & 1 & $2$\\
\hline\hline\hline
$a^*_{\mu}$ & 1 & -1 & 1 & 0 & $2$ \\
\hline
$c^*$ & 0 & -2 & 2 & 0 & $2$\\
\hline\hline
$\bar{c}^*$ & 0 & 0 & 0 & 0 & $2$\\
\hline
\end{tabular}
\end{center}
\caption{The various Abelian charges (\aka gradings) carried by the (anti)fields, 
namely Grassmann grading, 
ghost number, 
antighost/antifield number, pure gh \# = gh \# + ag \#, and mass dimension. 
The first two rows are the minimimal set of fields, the next two make it up to the non-minimal set, then the ensuing two rows are the minimal set of antifields, and $\bar{c}^*$ is needed for the non-minimal set.} 
\label{table:ghostantighost}
\end{table}

In table \ref{table:ghostantighost} we list the (anti)fields together with their  mass dimension, ghost number and Grassmann grading. These are assigned so that $S_0$ conserves these charges (\ie is bosonic and has zero dimension and ghost number) as will also be required of $S_I$. They imply that $\Delta$, $Q$ and $Q^-$ increase the dimension and ghost number by one. Similarly the antibracket \eqref{QMEbitsReg} adds one to the sum of the dimensions (sum of ghost charges) of $X$ and $Y$. 

We will see that in practice it is the free BRST cohomolgy that is central to solving for the effective action. The free BRST transformations follow from \eqref{BRS} or \eqref{Q0gen}, and \eqref{Sfreegimin}. The only non-vanishing action in  minimal basis is:\footnote{\label{foot:extended} In  extended gauge invariant basis \eqref{Sfreegi},
we would also have $Q_0\bar{c}=b$ and have a $\bar{c}^{(*)}$ part of $\Delta^-$.}
\be 
\label{Q0}
Q_0\, a_\mu = \partial_\mu c\,. 
\ee
while similarly from \eqref{KT}, 
the only non-vanishing actions of the free Kozsul-Tate operator are:
\be 
\label{Q0-}
Q^-_0 a^*_\mu = \Box\, a_\mu - \partial_\mu \partial\!\cdot\! a = \Box P^T_{\mu\nu} a_\nu\,,\qquad Q^-_0 c^* = - \partial\!\cdot\! a^*\,,
\ee
where $\Box = \partial_\mu^2$, and $P^T_{\mu\nu}$ is the transverse projector.
Following Henneaux \textit{et al}  \cite{Fisch:1989rp,Henneaux:1990rx,Barnich:1993vg,Barnich:1994db,Henneaux:1997bm,Gomez:2015bsa,Henneaux:1992ig}, we can simplify finding solutions of the free cohomology by splitting the problem up (\aka grading) by antighost (or depending on context, antifield) number. These are also listed in table \ref{table:ghostantighost}. They are chosen so that the free BRST charges have definite antighost number. We anticipated this with our labelling: $Q_0$  leaves antighost number unchanged, while $Q^-_0$ lowers it by one. Under this grading, the measure operator splits into two parts that lower antighost number by one or two respectively:
\be 
\label{Deltas}
\Delta = \Delta^- + \Delta^=\,,\qquad
 \Delta^- = -\frac{\partial}{\partial a^a_\mu} K \frac{\partial_r}{\partial a^{* a}_\mu}\,,\qquad
 \Delta^== \frac{\partial_r}{\partial c^a} K \frac{\partial}{\partial c^{* a}}\,.
\ee
The point of this extra grading is  that the action $S$ does not have definite anti-field number. We can therefore split it into parts of definite anti-field number $n$: 
$
S=\sum_{n=0} S^n\,.
$
To see why this is helpful, note that the full free BRST charge \eqref{fullBRS} can now be written
\be 
\label{fullBRSsplit}
\hat{s}_0 = Q_0+Q^-_0-\Delta^--\Delta^=\,.
\ee
This means that an operator $\Op = \sum_{m=0}^n \Op^m$ with some maximum antighost number $n$, that is {annihilated} by $\hat{s}_0$, must satisfy the descent equations:
\be 
\label{descendents}
Q_0\, \Op^n = 0\,, \quad Q_0\, \Op^{n-1} = (\Delta^--Q^-_0)\, \Op^n\,,\quad Q_0\,\Op^{n-2} =  (\Delta^--Q^-_0) \,\Op^{n-1} + \Delta^=\,\Op^n\,, \cdots\,.
\ee
Starting with the top (left-most) equation, these are often easier to analyse than trying to work with $\hat{s}_0\,\Op=0$ directly. Grading the square we also have the useful identities \cite{\morris}:
\beal
\label{nilpotents}
Q^2_0 =0\,,\ (Q^-_0)^2 = 0\,,\ &(\Delta^-)^2 = 0\,,\ (\Delta^=)^2 = 0\,,\nonumber\\
\{Q_0,Q^-_0\} =0\,,\ \{Q_0,\Delta^-\}=0\,,\ &\{Q^-_0,\Delta^=\}=0\,,\ \{\Delta^-,\Delta^=\}=0\,,\nonumber\\
\{Q^-_0,\Delta^-\}+& \{Q_0,\Delta^=\} = 0\,.
\eeal

\subsection{First order in coupling}
\label{sec:firstW}

Recall from \eqref{classicalCohomology} that we know that $S_{1,\text{cl}}$ is a local operator of dimension four which is a non-trivial solution of the full classical BRST cohomology. 
At this stage we could just write it down  and verify \eqref{classicalCohomology}.
 However it is more profound to notice, following \cite{Fisch:1989rp,Henneaux:1990rx,Barnich:1993vg,Barnich:1994db,Henneaux:1992ig,Henneaux:1997bm,Gomez:2015bsa}, that the solution is in fact unique up to {normaliz}ation and addition of a trivial reparametrisation $s_{0}\, \mathcal{K}_1$. 
 
 In fact for us even this latter term cannot appear since $\mathcal{K}_1$ would have to be an integrated local operator of ghost number $-1$, dimension three, and containing at least three (anti)fields. From table \ref{table:ghostantighost} no such operator exists.  For the same reasons we can see that the maximum possible antighost number of $S_{1,\text{cl}}$ is two. We can therefore write $S_{1,\text{cl}} = S^0_1 + S^1_1 + S^2_1$. If it is non-vanishing, this maximal term is unique up to {normaliz}ation (equivalent to {normaliz}ing $g$):
\be 
\label{S12}
S^2_1 = -i c^* c^2\,.
\ee 
(Any other arrangement of the (anti)fields is equivalent under cyclicity of the trace.) Now \eqref{classicalCohomology} implies that $S^2_1$ must satisfy
\be 
\label{top}
Q_0 S^2_1 = 0\,,
\ee
since this part alone has antighost number  two. Since from \eqref{Q0}, this is clearly true, it is a valid part of $S_1$ provided we can now find its descendants, $S^1_1$ and $S^0_1$. From \eqref{classicalCohomology} we must have that:
\be 
Q_0^- S^2_1 + Q_0 S^1_1 = 0\,.
\ee
Now by \eqref{Q0-}, 
\be 
Q_0^- S^2_1 = i \partial\!\cdot\! a^* c^2 = -i a^*_\mu \{ \partial_\mu c, c\} = -i a^*_\mu \{Q_0 a_\mu , c\} = i Q_0 \left( a^*_\mu [a_\mu,c] \right)\,.
\ee
where the second equality follows by integration by parts, $\{\,,\,\}$ being the anticommutator, and the third equality uses \eqref{Q0}. Thus, since cohomologically trivial parts are ruled out,
\be 
\label{S11}
S^1_1 = -i a^*_\mu [a_\mu,c] \,.
\ee
Finally, we must have $Q_0 S^0_1 = - Q^-_0 S^1_1$. 
Since
\beal
Q_0\left( -\frac{i}{2}(\partial_\mu a_\nu -\partial_\nu a_\mu) [a_\mu,a_\nu] \right) & = -i (\partial_\mu a_\nu -\partial_\nu a_\mu) [a_\mu,\partial_\nu c]\nonumber\\
&= i (\partial_\mu \partial\!\cdot\!a - \Box a_\mu) [a_\mu,c] = Q^-_0\left( i a^*_\mu [a_\mu,c] \right)\,,
\eeal
where in the second line we have integrated by parts and recognised that the resulting piece containing  $[\partial_\nu a_\mu,c]$ vanishes under the trace, we find
\be 
\label{S10}
S^0_1 = -i \partial_\mu a_\nu [a_\mu,a_\nu]\,.
\ee
Therefore we have verified conclusion \ref{point1} of sec. \ref{sec:five}. In fact the entire solution at first order in the coupling is:
\be 
\label{S1}
S_{1} = S_{1,\text{cl}} = -i c^* c^2 -i a^*_\mu [a_\mu,c] -i \partial_\mu a_\nu [a_\mu,a_\nu]\,,
\ee
since the quantum parts, being the tadpole quantum correction on the RHS of \eqref{flowSone}, and the measure operators in \eqref{fullBRS} and \eqref{Deltas}, give 
 vanishing contribution, in particular because they would leave behind a single field contracted into $\text{tr}\,T^a=0$.

\subsection{Classical solution at second order in coupling}
\label{sec:secondWC}

From \eqref{flow}, \eqref{flowSn} and \eqref{Sigman},
we need to  solve at second order at the classical level:
\beal 
\label{CME2general}
s_0 \,S_{2,\text{cl}} &= -\half (S_1,S_1)  \,,\\
\label{flow2}
\dot{S}_{2,\text{cl}} & = -\half\,\partial^r_A S_1 \dot{\propH}^{AB}\partial^l_BS_1  = -\half\, S_1\!\delr{A}\dot{\propH}^{AB}\dell{B}S_1\,
\eeal
(where $s_0$ was given in \eqref{classicalCohomology} and we also display a more intuitive notation for left and right derivatives \cite{Sonoda:2007dj,Igarashi:2009tj}). 

At first sight  we can solve for $S_{2,\text{cl}}$ already by using \eqref{CME2general}. Indeed from \eqref{QMEbitsReg}:\footnote{It is most straightforward to use $2\,\text{tr}(XT^a)\,\text{tr}(T^aY) = \text{tr}(XY)$ for traceless $X$, $Y$.}
\be 
\label{CME2}
s_0 \,S_{2,\text{cl}} = \left([c^{*},c] +[a_{\mu}^{*} ,a_{\mu}]\right) K (c^2) 
\,-\, \{a_{\mu}^{*}   ,c \} K [a_{\mu},c]
\,-\,  \Theta_\mu K [a_{\mu}, c]\,,
\ee
where $\Theta_\mu$ is shorthand for the result of differentiating \eqref{S10}:
\be 
\label{Theta}
\Theta_\mu := \partial_\rho[a_\rho,a_\mu] +[\partial_\mu a_\rho-\partial_\rho a_\mu,a_\rho]\,.
\ee
Counting antighost number, and bearing in mind that the $s_0$-closed part has already been solved for,
the highest antighost number that appears is $S^2_2\in S_{2,\text{cl}}$. Thus we know that:
\be 
\label{QS22}
Q_0\, S^2_2 = [c^*,c] K (c^2)\,.
\ee
If $K$ were not there, the right hand side would vanish on using the cyclicity of the overall trace. Therefore only pure derivative pieces survive in its derivative expansion:
\be 
K(-\partial^2/\Lambda^2) = 1 + \sum_{n=1}^\infty K_n (-\partial^2/\Lambda^2)^n\,.
\ee
Since any derivative acting on $c^2$ turns it into a $Q_0$-exact piece, and thus the whole of the right hand side into a $Q_0$-exact piece, we see that \eqref{QS22} has a solution.
However there is no natural canonical choice. Choosing different $\partial_\mu c$ to convert into $Q_0 \,a_\mu$,  corresponds to the freedom to add a piece that is trivially annihilated by $Q_0$, \ie to add 
an exact term $Q_0 \mathcal{K}^2_2 \in s_0 \mathcal{K}_2$. We see that now $\Lambda$ is involved, there are infinitely many solutions if we consider only \eqref{CME2general}. 


On the other hand, given that $S_1$ is $\Lambda$-independent,  the solution to \eqref{flow2} is uniquely determined by the requirement that it have a derivative expansion, up to addition of a $\Lambda$-independent dimension-four local operator $\Op_2$:
\be 
\label{S2cl}
{S}_{2,\text{cl}}  =  -\half\, S_1\!\delr{A}{\propH}^{AB}\dell{B}S_1 +\Op_2\,.
\ee
We recognise that the first part is the one-particle reducible (1PR) contribution: 
\be
\label{S2redDef} 
S_{2,red} = -\half\, S_1\!\delr{A}{\propH}^{AB}\dell{B}S_1\,,
\ee
whilst $\Op_2$ should also contain four (anti)fields but provide a 1PI (one-particle irreducible) contribution. The flow equation alone does not determine $\Op_2$, but we can already see from \eqref{CME2general}, or \eqref{CME2}, that if a solution exists for $\Op_2$ it is unique. This is because any change in $\Op_2$ would have to be $s_0$-closed. But we have already seen in sec. \ref{sec:firstW} that $S_1$ is the unique $s_0$-closed dimension four operator containing at least three (anti)fields.

Substituting \eqref{S1} into \eqref{S2redDef}, and using \eqref{Theta},
\eqref{propIR}, \eqref{propagators} and \eqref{gibasisDeriv}: 
\be 
\label{S2red}
S_{2,red} = \half \left( \{a^*_\mu,c\}+\Theta_\mu \right) \propH_{\mu\nu}\left( \{a^*_\nu,c\}+\Theta_\nu \right) \,-\, \left([c^*,c]+[a^*_\mu,a_\mu]\right) \propH\, \partial_\nu[a_\nu,c]\,. 
\ee
From this we see that 
\be 
Q_0\, S^2_{2,red} = - Q_0\, [c^*,c]\propH\, \partial_\nu[a_\nu,c] = [c^*,c]\propH \Box(c^2) = [c^*,c] K (c^2)\,,
\ee
and thus  that the 1PR part already satisfies the CME \eqref{CME2} at the antighost number two level, namely \eqref{QS22}. 

We also find that $S_{2,red}$ satisfies \eqref{CME2} at antighost number one. Indeed, recalling \eqref{Q0-}, defining $P^L_{\mu\nu} = p_\mu p_\nu/p^2$, and noting that
\be 
\label{QTheta}
Q_{0} \Theta_\mu = - \Box P^{T}_{\mu\nu}
[a_{\nu},c]\, + \,[ \Box P^{T}_{\mu\nu} a_{\nu},c]\,,
\ee
we verify the antighost number one descendent equation:
\beal
Q_{0}^{-} &\left( \half \{a_{\mu}^{*}  ,c \}\propH_{\mu\nu}\{a_{\nu}^{*}  ,c \} -[c^{*},c]\propH
\partial_\nu[a_{\nu},c]
\right)
 + Q_{0} \left( \Theta_\mu \propH_{\mu\nu}\{a^*_\nu,c\} -[a^*_\mu,a_\mu] \propH\, \partial_\nu[a_\nu,c] \right) \nonumber\\
&= -\{a_{\mu}^{*}  ,c \}\propH_{\mu\rho} \Box P^{T}_{\rho\nu}[a_{\nu},c] 
+ \{a_{\mu}^{*}  ,c \}\partial_\mu \propH\partial_\nu[a_\nu,c] - [a^*_\mu,a_\mu] \propH \Box (c^2)\nonumber\\
&= [a_{\mu}^{*} ,a_{\mu}] K c^2
\,-\, \{a_{\mu}^{*}   ,c \} K [a_{\mu},c]\,,
\eeal
where the first term on the second line comes from combining the first $Q^-_0$ and first $Q_0$ terms, and the next two terms from combining the second $Q^-_0$ and $Q_0$ terms. The final line coincides with the appropriate terms in \eqref{CME2}.

On the other hand, after some similar manipulation one finds the descendent equation at antighost number zero:
\be
Q^-_{0} \left( \Theta_\mu \propH_{\mu\nu}\{a^*_\nu,c\} -[a^*_\mu,a_\mu] \propH\, \partial_\nu[a_\nu,c] \right)
+ Q_0\left( \half \Theta_\mu \propH_{\mu\nu} \Theta_\nu\right) = -\Theta_\mu K [a_{\mu}, c] +[a_\mu,a_\nu][a_\mu,\partial_\nu c]\,,
\ee
and thus that $S_{2,red}$ violates the CME \eqref{CME2} by the last term on the right. At this point we recall that it is $S_{2,\text{cl}}$ that should satisfy the CME. From \eqref{S2cl}, that is therefore possible by choosing  
\be 
\label{O2}
\Op_2 = -\tfrac14 [a_\mu,a_\nu]^2\,.
\ee 
By the argument below \eqref{S2redDef}, this is the unique solution.
We recognise that it is the expected $g^2$ interaction in the classical bare action. And we note that we have now verified conclusion \ref{point2}. 

\subsection{Classical solution at third order in coupling}
\label{sec:thirdorderWC}


From \eqref{flow}, \eqref{flowSn} and \eqref{S2cl}, it is straightforward to see that the $g^3$ part of the classical solution can be integrated exactly. It is then uniquely determined by the requirement that it have a derivative expansion:
\be 
\label{S3clgeneral}
S_{3,\text{cl}} = -\, S_1\!\delr{A}\!\propH^{AB} \dell{B}\Op_2\ +\ \half\, S_1\!\delr{A}\propH^{AB} \left(\dell{B}S_1\!\delr{C}\right) \propH^{CD} \dell{D}S_1\,,
\ee
since from the argument of  \ref{point3}, 
by dimensions
there can be no integration constant. In particular in our case, the $S_{n,\text{cl}}$ have precisely $n\cu+2$ (anti)fields.
In sec. \ref{sec:legTrans} we will also derive \eqref{S3clgeneral} from the Legendre transform identity \cite{Morris:1993}. 

Recall from sec. \ref{sec:five} that similarly, thanks to compatibility with the flow equation,  the $g^3$ part of the CME has only the potential to be violated by a $\Lambda$-independent local term. Since such a term must have ghost number one and dimension five, it must contain precisely four $a$ fields and one $c$. Then given also that it must be single trace and valued in the Lie algebra (be composed of the product of commutators) one can show that such terms vanish identically. We will see this explicitly at the end of this section.

Therefore we must already find:
\be 
\label{Sigmac3}
\Sigma_{3,\text{cl}} = s_0 \,S_{3,\text{cl}} + (S_1,S_{2,\text{cl}}) = 0\,.
\ee
Now, substituting for $\Op_2$ and $S_1$ in \eqref{S3clgeneral}, using \eqref{O2} and \eqref{S1}, we find that:
\beal 
S_{3,\text{cl}} 
= &-i\, (\{a_{\mu}^{*},c \} + \Theta_{\mu})\,
\propH_{\mu\nu}\bigl[[a_{\nu},a_{\sigma}],a_{\sigma}\bigr]  \nn\\
&- i\bigl([c^{*},c] +  [a_{\kappa}^{*},a_{\kappa} ] \bigr) 
\Bigl(-\bigl[(\pa \propH) \cu\cdot a\,
 \propH, \pa \cu\cdot [a,c] \bigr]
+ \bigl[ \bigl(\pa_{\mu}\propH \bigr)c 
\,\propH_{\mu\nu},
\{a^{*}_{\nu},c\} + \Theta_{\nu}
\bigr] 
\Bigr) \nn\\
&- i \bigl(\{a^{*}_{\mu},c\} + \Theta_{\mu}
\bigr)\bigl\{\propH_{\mu\nu}\,a_{\nu}^{*}
\,\propH, \pa \cu\cdot [a,c] \bigr\}
\nn\\
&
+ \tfrac{i}2  \bigl(\{a^{*}_{\mu},c\} + \Theta_{\mu}
\bigr)
\Bigl[\bigl(\pa_{\rho}\propH_{\mu\nu}\bigr)\,
a_{\nu}\,\propH_{\rho\sigma} - \bigl(\pa 
\propH_{\mu\nu}\bigr)\cu\cdot a \,
\propH_{\nu\sigma} \nn\\
&
+ \propH_{\mu\nu} \Bigl(a_{\rho} \bigl(\pa_{\nu} \propH_{\rho\sigma}\bigr) 
- \bigl(\pa_{\nu}a_{\rho}\bigr)\propH_{\rho\sigma}- 
a \cu\cdot \bigl(\pa \propH_{\nu\sigma}\bigr) + \bigl(\pa_{\rho}
a_{\nu}\bigr)\propH_{\rho\sigma} \Bigr),
\{a_{\sigma}^{*},c \} + \Theta_{\sigma}\Bigr]
\nn\\
&
- \tfrac{i}{2} \pa \cu\cdot [a,c] \bigl[\propH\,
c^{*} \propH,\pa \cu\cdot [a,c]\bigr]\,,
\label{S31}
\eeal
where in the above equation it should be understood that the kernels ($\propH$ and $\partial\propH$) do not take part in the (anti)commutators (see app. \ref{app:notation} for more explanation). 
We see that all terms in $\Sigma_{3,\text{cl}}$ except $(S_{1},\Op_2)$ are 1PR. Even $(S_{1},\Op_2)$ is non-local however, since it contains the regulator $K$. From \eqref{S2red}, the 
antibracket $(S_{1},S_{2,red})$ consists of terms with a factor $K$ and one (ghost or gauge) propagator. These terms collect together with parts of $s_0\, S_{3,\text{cl}}$ in such a way as to eliminate one propagator from $s_0 S_{3,\text{cl}}$. Splitting by antighost number, we thus find:
\beal
&
\Sigma_{3,\text{cl}}^{2} = i [c^{*},c] 
\Bigl(\bigl\{c\, \pa_{\mu}\propH ,[a_{\mu},c]
\bigr\} -\bigl\{ \bigl(\pa_{\mu}
  \propH \bigr)c,[a_{\mu},c]\bigr\}
+ \bigl[\bigl(\pa_{\mu} \propH\bigr) a_{\mu},
\frac{1}{2}\{c,c\} \bigr]\Bigr)
\label{Sigma3cl12}\\
& 
\qquad+ i [a_{\mu},c] \bigl[(\pa_{\mu}\propH) c^{*}, 
\frac{1}{2}\{c,c\} \bigr]
+ i \{a_{\mu}^{*},c \}\Bigl(\bigl[ c\, \propH_{\mu\nu},\{a_{\nu}^{*},c \}
\bigr] 
- \bigl[\propH_{\mu\nu}\,a_{\nu}^{*}, \frac{1}{2}\{c,c\}\bigr]
 \Bigr)
\nn\\
& 
\Sigma_{3,\text{cl}}^{1} = i [a_{\kappa}^{*},a_{\kappa} ]\bigl\{c\,
\propH ,\pa \cu\cdot [a,c]\bigr\} 
- i \{a^{*}_{\mu},c\}
\Bigl(
 \propH_{\mu\nu}\bigl(
\bigl[[\pa_{\nu}c,a_{\sigma}],a_{\sigma}\bigr] 
+ \bigl[[a_{\nu},
  \pa_{\sigma}c],a_{\sigma}\bigr] 
+ \bigl[[a_{\nu},a_{\sigma}],\pa_{\sigma}c
  \bigr]\bigr)
\nn\\
& 
\quad-
\bigl[\pa_{\sigma}\propH_{\mu\nu} 
a_{\nu} - \pa \propH_{\mu\sigma} \cu\cdot a 
+ \propH_{\mu\nu}(\pa_{\sigma} a_{\nu} - \pa_{\nu} a_{\sigma}) 
,[a_{\sigma},c]\bigr]  
- \bigl[\bigl(\propH_{\mu\nu} \,a_{\sigma}
- \propH_{\mu\sigma}\, a_{\nu}\bigr) ,\pa_{\nu}[a_{\sigma},c]\bigr]
\nn\\
& 
+\bigl[a_{\mu} \propH,\pa \cu\cdot [a,c] \bigr] - \bigl[c \propH_{\mu\nu}, \Theta_{\nu}
\bigr] + \propH_{\mu\nu}\bigl[c,\Theta_{\nu}\bigr] 
\Bigr) + i c^{2}  \bigl[a^{*}_{\mu}\propH_{\mu\nu},
\Theta_{\nu}
\bigr] -i [a_{\mu},c] \bigl\{a_{\mu}^{*} \propH, \pa\cu\cdot [a,c]\bigr\} 
\nn
\eeal
Using trace properties such as the Jacobi identity, 
we can then show that $\Sigma_{3,\text{cl}}^{2} = \Sigma_{3,\text{cl}}^{1} =0$. 
Finally for antighost number 0, 
\beal
\Sigma_{3,\text{cl}}^{0} &= -i [a_{\mu},c] K \bigl[a_{\nu},[a_{\nu},a_{\mu}]\bigr] 
- i\Theta_{\mu} 
\bigl[a_{\mu} \propH,\pa \cu\cdot [a,c] \bigr] 
-i \bigl( - \Box P^{T}_{\mu\rho}
([a_{\rho},c]) \bigr)
   \propH_{\mu\nu}\bigl[[a_{\nu},a_{\sigma}],a_{\sigma}\bigr]
\nn\\ 
&= -i[a_{\mu},c] \Bigl(K \bigl[a_{\nu},[a_{\nu},a_{\mu}]\bigr]
+ (\delta_{\mu\nu}(1-K) + \pa_{\mu}\pa_{\nu}\propH)
\bigl[a_{\sigma},[a_{\sigma},a_{\nu}]\bigr]  
\nn\\
 &\qquad
+ \pa_{\mu}\propH \bigl[a_{\rho}, \pa_{\kappa}[a_{\kappa}, a_{\rho}]
+ [\pa_{\rho}a_{\kappa},a_{\kappa}] + [a_{\kappa},\pa_{\kappa}a_{\rho}]\bigr]
\Bigr) = -i[a_{\mu},c]\bigl[a_{\nu},[a_{\nu},a_{\mu}]\bigr] =0 \,.
\label{sigma30}
\eeal 
Note that the first term in the first line of (\ref{sigma30}) is $(S_{1},\Op_2)$. Its $K$ part gets eliminated by a similar $1\cu-K$ piece  leaving behind the local term in the last line which is of the form that we flagged at the beginning of this section, and thus vanishes.\footnote{For example under the trace it is the same as
$
i\left[a_\nu,[a_{\mu},c]\right] [a_{\nu},a_{\mu}] = \frac{i}2\left[ [a_\nu,a_\mu],c\right] [a_{\nu},a_{\mu}] =\frac{i}4\left[ [a_\nu,a_\mu]^2,c\right] = 0
$.} It corresponds to the unregularised version of $(S_{1},\Op_2)$ (and also coincides with the term that appears in the 1PI formulation in sec. \ref{sec:secondGC}),
and thus its vanishing is the statement that the bare classical Yang-Mills action has no $g^3$ interaction.

We have thus for this example, verified conclusion \ref{point3}. 

\subsection{One-loop solution at second order in coupling}
\label{sec:secondorderWloop}

We now turn to the one-loop corrections. We have already seen at the end of sec. \ref{sec:firstW}, that $S_1$ is purely classical. On the other hand $S_2$ has a one-loop part, $S_{2,\text{q}}$. From \eqref{flowSn}, \eqref{Sigman} and the definition of $s_0$ in \eqref{classicalCohomology}, it satisfies:
\be
\label{S2qRelations}
\dot{S}_{2,\text{q}} = -a_1[S_{2,\text{cl}}] 
\qquad\text{and}\qquad
s_0 S_{2,\text{q}} = \Delta S_{2,\text{cl}}\,.
\ee
From \eqref{flow} and the explicit formulae for $S_{2,\text{cl}}$, \viz \eqref{S2red} and \eqref{O2}, and recalling the rule \eqref{gibasisDeriv} for computing in gauge invariant basis, we see that $S_{2,\text{q}}$ is made of the two-point vertices:\footnote{Strictly we should write $\mathcal{A}_{\mu\nu}(i\partial)$ if we regard it as defined by its momentum representation as we do for $\mathcal{B}$.}
\be 
\label{S2q}
S_{2,\text{q}} =   \half\,C_A\, a_\mu\, \mathcal{A}_{\mu\nu}(\partial)\, a_\nu + C_A\, a^*_\mu\, \mathcal{B}(-\partial^2)\, \partial_\mu c  \,,
\ee
where $ C_A\,\mathcal{A}_{\mu\nu}(p)$ and $C_A\,\mathcal{B}(p^2)\, p_\mu$ are the $\Lambda$-integrals of one-loop Feynman diagrams 
formed from  attaching $\dot{\propH}$ for the ghosts, and also for the gauge-field.
 For later convenience we have pulled out a factor of $C_A$, where $C_A \,\delta^{ab} = f^{acd}f^{bcd}$ defines the Casimir of the adjoint representation, and for $\mathcal{B}$ we have used Lorentz invariance to pull out a factor of the external momentum $p_\mu$. 
The $\Lambda$-integrals can in fact be done exactly, as will become especially clear when we consider just the flow of the 1PI part. We therefore 
reserve their derivation until secs. \ref{sec:oneloopG} and \ref{sec:oneloopsecondorder}.

From \eqref{Deltas} we see that $\Delta S_{2,\text{cl}}$ results in one-loop diagrams formed by the action of $\Delta^{-}$ and $\Delta^{=}$ on the antighost-number one and two parts respectively of \eqref{S2red}, and thus we find:
\be 
\Delta S_{2,\text{cl}} = C_A\, a_\mu\, \mathcal{F}(-\partial^2)\,\partial_\mu c \,,
\ee
where 
\beal 
\mathcal{F}(p^2) &= \frac{p_\mu}{p^2} \int_q \propH(q) K(p\!+\!q)
  \Bigl[(3 + \xi) q_{\mu} + (2 + \xi) p_{\mu} + (1-\xi) 
{\frac{q_{\mu}}{q^{2}}}\,(p+q)\cu\cdot q \Bigr]\nn\\
& =\int_q \propH(q) K(p\!+\!q) \left\{ 4 \frac{p\!\cdot\!q}{p^2}+2+\xi+(1-\xi)\frac{(p\!\cdot\!q)^2}{p^2q^2}\right\}\,.
\label{F}
\eeal
In the first line we exploit the fact that the Feynman integral is proportional to $p_\mu$. Our notation is:
\be 
\label{momentumNotation}
a_\mu(x) = \int_p a_\mu(p) \,\text{e}^{-ip\cdot x}\,,\qquad \int_p \equiv \int \frac{d^4p}{(2\pi)^4}\,.
\ee
Now from \eqref{S2q}, since $s_0=Q_0+Q^-_0$ where the latter are given in \eqref{Q0} and \eqref{Q0-}, we see that the second equation in \eqref{S2qRelations} implies no constraint on $\mathcal{B}$ but
\be 
\label{AWI}
p_\mu\,\mathcal{A}_{\mu\nu}(p) =  \mathcal{F}(p^2)\, p_\nu\,.
\ee
From \ref{point4}, 
this should be already satisfied, up to a possible $\Lambda$-independent piece in $\mathcal{A}_{\mu\nu}(p)$. 
In sec. \ref{sec:oneloopsecondorder} we show that 
for the natural form of $\mathcal{A}_{\mu\nu}(p)$,
\eqref{AWI} is in fact automatically satisfied.

\section{The Legendre flow equation and the mST}
\label{sec:leg}

The 1PI parts of the Wilsonian effective action $S$ are in fact the vertices of the Legendre effective action ($\Gamma^{tot}$) regulated by the infrared cutoff $\bar{K}$ \cite{Morris:1993,Ellwanger1994a,Rosten:2010pc,Morris:2015oca,Bonini:1992vh,Keller:1990ej}:
\be 
\label{Gammatot}
\Gamma^{tot}[\ph,\ph^*] = \half\, \ph^A \propH^{-1}_{AB}\ph^B + \ph^*_A R^A_{\ B} \ph^B +\Gamma_I[\Phi,\Phi^*]\,.
\ee
In other words, $S$ can be computed through a tree-level expansion, using vertices from $\Gamma_I$ \cite{Morris:1993,Morris:2015oca}. It is therefore most efficient to solve first for these 1PI parts. In secs. \ref{sec:firstG} -- \ref{sec:oneloopG}, we recover the previous results this way, and then in sec. \ref{sec:oneloopsols} we evaluate the one-loop contribution and also take it further, to $O(g^3)$. 

First we demonstrate by means of the Legendre transformation \cite{Morris:1993,Ellwanger1994a,Rosten:2010pc,Morris:2015oca} that $\Gamma_I$ is governed by the flow of the effective average action \cite{Morris:1993,Nicoll1977,Weinberg:1976xy,Morris:2015oca,Bonini:1992vh,Morris:1995he,Morgan1991,Wetterich:1992,Ellwanger1994a} and the mST identities \cite{Ellwanger:1994iz}. In secs. \ref{sec:firstG} -- \ref{sec:oneloopG} we also see how the two formulations match together.
In sec. \ref{sec:general} we demonstrate that the freedom in the continuum solution is parametrised by wavefunction renormalization constants that generate a simple canonical transformation and solve the standard Slavnov-Taylor identities.

\subsection{Legendre transform of the Wilsonian flow equation and QME}
\label{sec:legTrans}

In the above equation \eqref{Gammatot}, $\Phi^*=\phi^*$ are the same antifield sources, renamed for aesthetics, $\propH^{-1}_{ AB}$ is the inverse of the IR regulated propagator \eqref{propIR}, and $\Phi^A= \{ A,B,C,\bar{C}\}$ are the so-called classical fields. The interaction parts of each effective action are related via the Legendre transform relation \cite{Morris:1993,Ellwanger1994a,Rosten:2010pc,Morris:2015oca}:
\be 
\label{Legendre}
\Gamma_I[\ph,\ph^*] = S_I[\Ph,\Ph^*] - \half\, (\Ph -\ph)^A\, \propH^{-1}_{ AB}\,(\Ph-\ph)^B\,.
\ee
This means in particular that
\be 
\label{LegendreIdentities}
\frac{\partial_r}{\partial\phi^B} S_I[\phi,\phi^*] =  (\phi-\Phi)^A \propH^{-1}_{AB} = \frac{\partial_r}{\partial\Phi^B} \Gamma_I[\Phi,\Phi^*]\,,
\ee
and similarly for left derivatives, thus substituting back we can write \cite{Morris:1993,Morris:2015oca}:
\be
\label{halfway}
S_I[\Ph,\Ph^*] = \Gamma_I[\phi-\propH\partial^l S_I,\phi^*] +\half\,\partial^r_A S_I\propH^{AB}\partial^l_{B}S_I\,.
\ee
Taylor expanding the $\Gamma_I$ term, and iteratively substituting for $S_I[\phi,\phi^*]$ then gives the desired tree (1PR) expansion:
\beal
S_I & = \Gamma_I[\phi,\phi^*] -\Gamma_I\!\delr{A}\!\propH^{AB}\!\dell{B}\! S_I + \half\, S_I\!\delr{A}\!\propH^{AB}\!\dell{B}\!\Gamma_I\!\!\delr{C}\!\propH^{CD}\!\dell{D}\!S_I +\cdots+\half\,S_I\!\delr{A}\!\propH^{AB}\!\dell{B}\!S_I\,,\nn\\
&= \Gamma_I[\phi,\phi^*] -\half\,\Gamma_I\!\delr{A}\!\propH^{AB}\!\dell{B}\! \Gamma_I + \half\, \Gamma_I\!\delr{A}\!\propH^{AB}\!\left(\dell{B}\!\Gamma_I\!\!\delr{C}\right)\!\propH^{CD}\!\dell{D}\!\Gamma_I + O\left(\Gamma^{\ 4}_I\right)\,.
\label{treeExpansion}
\eeal
For example, specialising to the purely classical part reproduces the general form of the solutions \eqref{S2cl} and \eqref{S3clgeneral}, on identifying $\Gamma_{I,\text{cl}}$ with the 1PI part of $S_{I,\text{cl}}$. 

We now use the Legendre transformation to confirm the form of the flow equation and modified Slavnov Taylor (mST) identities for $\Gamma_I$. Indeed from \eqref{Legendre} and \eqref{LegendreIdentities}, 
\be 
\partial_t|_\Phi \Gamma_I = \partial_t|_\phi S_I + \half\, (\Ph -\ph)^A \left(\propH^{-1} \dot{\propH}\propH^{-1}\right)_{ AB}\!(\Ph-\ph)^B = \dot{S}_I + \frac12\frac{\partial_r S_I}{\partial\phi^A} \dot{\propH}^{AB}\frac{\partial_lS_I}{\partial\phi^B}\,.
\ee
Using \eqref{flow}, and differentiating \eqref{LegendreIdentities} once more with respect to the fields, one has:
\be 
\label{inverseGammaOp}
\frac{\partial_r\Phi^A}{\partial\phi^B} = \left(\left[1+\propH \Gamma^{(2)}_I \right]^{-1}\right)^{\!A}_{\ \ B} \,,
\ee
and thus we derive the flow equation for $\Gamma_I$ \cite{Morris:1993}:
\be 
\label{GammaFlow}
\dot{\Gamma}_I = -\half\, \text{Str}\left( \dot{\propH}\propH^{-1} \left[1+\propH \Gamma^{(2)}_I \right]^{-1}\right) 
\,,
\ee
see also \cite{Nicoll1977,Weinberg:1976xy,Morris:2015oca,Bonini:1992vh,Morris:1995he,Morgan1991},  where Str$\,\mathcal{M} = (-)^A\, \mathcal{M}^{A}_{\ \,A}$, and
we have introduced
\be 
\label{Hessian}
\Gamma^{(2)}_{I\ AB} 
=  \frac{\dell{}}{\partial\Phi^A}\Gamma_I\frac{\delr{}}{\partial\Phi^B}\,.
\ee

It is a short step to recast this as the flow for the effective average action \cite{Wetterich:1992,Ellwanger1994a}, $\Gamma$, but it is \eqref{GammaFlow} that will be useful here.
On the other hand, the mST is best expressed in terms of  $\Gamma$.  This 1PI effective action is just the {IR-cutoff} Legendre effective action after subtracting the infrared cutoff term, where the latter is expressed in additive form:
\be 
\label{LegendreEffAct}
\Gamma^{tot} = \Gamma + \half \ph^A \mathcal{R}_{AB} \ph^B\,,  \qquad \propH^{-1}_{ AB} = \prop^{-1}_{AB} + \mathcal{R}_{AB}\,.
\ee
As implied by \eqref{Gammatot}, $\Gamma$ is thus expressed in terms of a free part, $\Gamma_0$, that carries no regularisation:
\be 
\label{Gamma}
\Gamma = \Gamma_0 + \Gamma_I\,,\qquad \Gamma_0 =
\half\, \ph^A \prop^{-1}_{AB}\ph^B + \ph^*_A R^A_{\ B} \ph^B\,.
\ee
Consistent with this, the antibracket 
is most naturally expressed without regularisation. Thus for arbitrary functionals of the classical (anti)fields, $\Xi[\Phi,\Phi^*]$ and $\Upsilon[\Phi,\Phi^*]$, we define
\be 
\label{QMEbitsPhi}
(\Xi,\Upsilon) = \frac{\partial_r\Xi}{\partial\Phi^A}\,\frac{\partial_l\Upsilon}{\partial\Phi^*_A}-\frac{\partial_r\Xi}{\partial\Phi^*_A}\,\frac{\partial_l\Upsilon}{\partial\Phi^A}\,.
\ee
To arrive at the mST, we also need from \eqref{Legendre} that (and similarly for left derivatives):
\be 
\label{LegendreAntiIdentities}
\frac{\partial_r}{\partial\phi^*_A}\Big|_\phi S_I[\phi,\phi^*] = \frac{\partial_r}{\partial\Phi^*_A}\Big|_\Phi \Gamma_I[\Phi,\Phi^*]\,.
\ee
Recalling the $K$ regularisation in the antibracket \eqref{QMEbitsReg}, and using 
\eqref{S0}, we have 
\beal
\half(S,S) &= (S_I,S_0) +\half(S_I,S_I)\nn\\ 
&= \frac{\partial_r S_I}{\partial\Ph^A} R^A_{\ B}\Ph^B -\frac{\partial_r S_I}{\partial\Ph^*_A} \left(\prop^{-1}_{AB}\Ph^B+(-)^A R^B_{\ A} \Ph^*_B\right) + \frac{\partial_r S_I}{\partial \Ph^*_A}\bar{K} \frac{\partial_l S_I}{\partial \Ph^A}
-\frac{\partial_r S_I}{\partial \Ph^*_A} \frac{\partial_l S_I}{\partial \Ph^A} \nn\\
&= \frac{\partial_r \Gamma_I}{\partial\ph^A} R^A_{\ B} \left(\ph^B+ \frac{\partial_r\Gamma_I}{\partial\ph^C}\propH^{CB}\right) -\frac{\partial_r \Gamma_I}{\partial\ph^*_A} \left(\prop^{-1}_{AB}\Phi^B+(-)^A R^B_{\ A} \Phi^*_B\right) +\half(\Gamma_I,\Gamma_I)\nn\\
&= (\Gamma_I,\Gamma_0)+ \half(\Gamma_I,\Gamma_I) = \half (\Gamma,\Gamma)\,,
\label{CMEequivalence}
\eeal
where in the third line we used \eqref{LegendreIdentities} and \eqref{LegendreAntiIdentities}, in particular converting $\bar{K}\partial^l_A S_I$ to $\prop^{-1}_{AB}(\phi-\Phi)^B$, and the antibracket \eqref{QMEbitsPhi}. Then  we note that the second term in the first bracket vanishes because 
\be 
\label{linearisedIdentity}
\prop^{CB}R^A_{\ B} + \prop^{AB}R^C_{\ B} = 0\,,
\ee
by linearised BRST invariance (see app. \ref{app:leftright}), after which we are left with the expanded version of $(\Gamma_I,\Gamma_0)$, as one can see by comparing to the analogous terms in the second line. Finally, 
\be 
\Delta S_I =  \frac{\partial_r}{\partial\phi^A}\,K\frac{\partial_lS_I }{\partial\phi^*_A} 
= \frac{\partial_r}{\partial\Phi^B}\left(\!K \frac{\partial_l\Gamma_I}{\partial\Phi^*_A}  \right) \frac{\partial_r\Phi^B}{\partial\phi^A} =  \text{Tr}\left( \!K  \Gamma^{(2)}_{I*} \left[1+\propH\Gamma^{(2)}_I\right]^{-1}\right)\,,
\ee
where we used \eqref{QMEbitsReg} and the fact that $S_I$ is bosonic, followed by \eqref{LegendreAntiIdentities} then \eqref{inverseGammaOp}, set
\be 
\label{HessianStar}
\left(\Gamma^{(2)}_{I*}\right)^{A}_{\ \ B} 
\,=\,  \frac{\dell{}}{\partial\Phi^*_A}\Gamma_I\frac{\delr{}}{\partial\Phi^B}\,,
\ee
and write Tr$\,\mathcal{M} = \mathcal{M}^{A}_{\ \,A}$.
Thus we have shown that the QMF \eqref{QMF} can be written as
\be 
\label{mST}
\Sigma = \half (\Gamma,\Gamma) - \text{Tr}\left( \!K  \Gamma^{(2)}_{I*} \left[1+\propH\Gamma^{(2)}_I\right]^{-1}\right)\,.
\ee
Its vanishing is the QME. We see that it coincides with the mST identities first derived in ref. \cite{Ellwanger:1994iz}.


\subsection{Discussion}
\label{sec:discussion}

In the limit $\Lambda\to0$ (holding (anti)fields fixed), we have that $\Gamma\to\Gamma^{tot}$, the standard Legendre effective action without cutoff, as is clear from \eqref{Gammatot}. The correction terms in \eqref{mST} are  UV regularised by $K$ (and IR regularised by $\bar{K}$). They therefore vanish in this limit. The antibracket \eqref{QMEbitsPhi} is not regularised and thus the mST tends to the Zinn-Justin equation \cite{ZinnJustin:1974mc,ZinnJustin:1975wb,ZinnJustin:2002ru,Ellwanger:1994iz}, \aka CME,
\be 
\label{limitisZJ}
0=\Sigma\to \half(\Gamma,\Gamma)\qquad\text{as}\qquad \Lambda\to0\,,
\ee
as desired. As we will see in sec. \ref{sec:firstG}, the mST and this CME lead to analogous definitions for the BRST charges and Koszul-Tate operator.

Although the antibracket is not regularised, it only introduces trees. Then, as we will see, UV regularisation provided in the correction terms in \eqref{mST}, and through \eqref{GammaFlow}, is enough to ensure that expressions are well defined. Introducing an overall UV regularisation  $K_0= K(p^2/\Lambda_0^2)$, by writing $1-K\mapsto K_0 -K$ in \eqref{Legendre} \cite{Morris:1993,Morris:2015oca}, does not change the form of the flow equation \eqref{GammaFlow}. However the regularised propagator becomes $\propH = \prop (K_0-K)$. Following through the derivations, it indeed makes its appearance this way in \eqref{mST}, while the antibracket \eqref{QMEbitsPhi} is then regularised by $K_0$.

The antibracket $(\ ,\ )_\phi$ for quantum fields introduced in \eqref{QMEbitsReg}, implies the canonical structure $(\phi^A,\phi^*_B)_\phi = K\delta^A_B$. However for this antibracket, $\Phi$ is not canonically conjugate to $\Phi^*$: 
\be
(\Phi^A,\Phi^*_B)_\phi = (\Phi^A,\phi^*_B)_\phi = K\! \frac{\partial_r}{\partial\phi^B}\, \Phi^A \ne K\delta^A_B\,.
\ee
Indeed, the antibracket $(\ ,\ )_\Phi$ introduced in \eqref{QMEbitsPhi} is not canonically related to $(\ ,\ )_\phi$. Although $\Sigma$ is invariant under quantum canonical transformations $S\mapsto S+\varepsilon \hs\mathcal{K}$, \cf \eqref{sK}, the correction term in the mST \eqref{mST} provides an obstruction to canonical transformations of $\Gamma$ that respect $(\ ,\ )_\Phi$.
In particular the analogous canonical transformation to \eqref{gfgi}, namely
\beal
\bar{C}^*  |_\text{gf}  &= \bar{C}^*  |_\text{gi} - i \partial\!\cdot\! A\,, \nonumber\\
A^*_\mu |_\text{gf} & = A^*_\mu |_\text{gi} + i \partial_\mu \bar{C}\,,
\label{GFGI}
\eeal
does not leave \eqref{mST} invariant. However, since \eqref{gfgi} does leave $\Sigma$ invariant, the Legendre transformation \eqref{Legendre} implies that the mST must be invariant under a non-linear transformation, as we show in app. \ref{app:mSTtrans}.

On the other hand, as we will confirm in sec. \ref{sec:firstG}, we can still use the arguments at the end of sec. \ref{sec:S0} to see that $\Gamma_I$ does not depend on $\bar{C}^*$ or $B$, and depends on $\bar{C}$ only through the combination $A^*_\mu - i \partial_\mu\bar{C}$. Then we can work, as we will do from now on, in minimal gauge invariant basis, provided we make the replacement (or equivalently temporarily replace $A^*_\mu$ with $A^*_\mu - i \partial_\mu\bar{C}$)
\be 
\label{GIbasisDeriv}
\frac{\partial}{\partial\bar{C}}  \Big |_\text{gf}\ \equiv\   i\partial_\mu\frac{\partial}{\partial{A}^*_\mu} \Big |_\text{gi}
\ee
in both the flow equation \eqref{GammaFlow} and the mST \eqref{mST}.

For example, in this basis we can write the free action $\Gamma_0$ explicitly as:
\be 
\label{sol0}
\Gamma_0 = \half \left( \partial_\mu A_\nu \right)^2  -\half \left(\partial\!\cdot\! A\right)^2   + A^*_\mu \partial_\mu C\,,
\ee
as can be seen by comparing \eqref{Gamma}, \eqref{S0} and \eqref{Sfreegimin}.
In the next sections, we show how perturbative expansion of the flow \eqref{GammaFlow} and mST \eqref{mST} via:
\be 
\label{Gexpand}
\Gamma = \Gamma_0 + g\Gamma_1 + g^2\Gamma_2 + g^3 \Gamma_3 +\cdots\,, 
\ee
rephrases the properties and results for the Wilsonian action in secs. \ref{sec:firstW} -- \ref{sec:secondorderWloop}. 

\subsection{Perturbations: first order in coupling}
\label{sec:firstG}

Expanding  \eqref{mST} to first order, we get the same expression for the free full BRST charge in terms of the classical fields:
\be 
\label{mST1}
\hat{s}_0\, \Gamma_1 = (Q_0+Q^-_0-\Delta^--\Delta^=)\, \Gamma_1 = 0\,.
\ee
Analogous to \eqref{BRS} and \eqref{KT}, we have defined the right-acting
\be 
Q\,\Phi^A = (\Phi^A,\Gamma) = \frac{\partial_l\Gamma}{\partial\Phi^*_A}\qquad\text{and}\qquad Q^-\Phi^*_A = (\Phi_A^*,\Gamma)=-\frac{\partial_l\Gamma}{\partial\Phi^A}\,,
\ee
the full quantum BRST charge, and full classical BRST charge:
\be \hs\,\Op = (Q+Q^--\Delta) \Op\,, \qquad s_\text{cl}\,\Op = (Q_\text{cl}+Q^-_\text{cl})\, \Op = (\Op,\Gamma_\text{cl})\,. \label{fullBRSTG}\ee
This leads to the same free algebra as \eqref{Q0} and \eqref{Q0-}, which in particular are again  unregulated: 
\be 
\label{BRSfree}
Q_0\, A_\mu = \partial_\mu C\,,\qquad Q^-_0 A^*_\mu = \Box\, A_\mu - \partial_\mu \partial\!\cdot\! A\,,\qquad Q^-_0 C^* = - \partial\!\cdot\! A^*\,,
\ee
but this time because neither $\Gamma_0$ nor the antibracket is regulated. On the other hand the measure operator appears from the first order part of \eqref{mST} as 
\be
\Delta \Gamma = \text{Tr}\left( K  \Gamma^{(2)}_{*} \right)\,,
\ee
and thus $\Delta = \Delta^- + \Delta^=$ also continues to take the same form: 
\be 
\label{DeltasG}
 \Delta^- = \frac{\partial}{\partial A^a_\mu} K \frac{\partial_l}{\partial A^{* a}_\mu}\,,\qquad \Delta^== - \frac{\partial_l}{\partial C^a} K \frac{\partial}{\partial C^{* a}}\,,
\ee 
in particular continues to be regulated by $K$. 

Since \eqref{mST1} is thus identical in form to the analogous equation for $S_1$, and since at the classical level \eqref{GammaFlow} just says that the operator must be $\Lambda$ independent,
we have the same unique (up to {normaliz}ation) non-trivial solution:
\be 
\label{sol1}
\Gamma_1 = \Gamma_{1,\text{cl}} =
-i C^*C^2 -iA^*_\mu [A_\mu,C] -i \partial_\mu A_\nu [A_\mu,A_\nu]\,. 
\ee
This is trivially a first order solution at the quantum level also, for the same reasons as before. Indeed, \eqref{DeltasG} takes the same form, as does \eqref{GammaFlow} to this order:
\be 
\label{flow1}
\dot{\Gamma}_1 = \half\, \text{Str}\left( \dot{\propH}\Gamma^{(2)}_1\right) = -a_1[\Gamma_1]\,,
\ee
by \eqref{flow}.
Thus we have constructed the cohomologically unique first order simultaneous solution of the mST and flow equation for the 1PI effective action.

\subsection{Higher order in coupling at the classical level}
\label{sec:secondGC}

At the classical level, the flow equation \eqref{GammaFlow} just determines $\Lambda$-integration constants, since it says that $\Gamma_{\text{cl}}$ must be $\Lambda$ independent. Meanwhile the mST \eqref{mST} reduces to the CME. Thus the simultaneous solution is just the same as obtained in classical BRST cohomology for a local dimension four operator.  

At $O(g^2)$:
\be 
s_0\, \Gamma_{2,\text{cl}} = (Q_0+Q_0^-)\, \Gamma_{2,\text{cl}} = -\half (\Gamma_1,\Gamma_1) = Q_0 \left(-\tfrac14[A_\mu,A_\nu]^2\right)\,,
\ee 
where we substituted \eqref{sol1}. This provides us directly with the unique classical integration constant:
\be 
\label{sol2c}
\Gamma_{2,\text{cl}} = -\tfrac14[A_\mu,A_\nu]^2\,.
\ee
We note that although  this coincides with the (1PI) integration constant \eqref{O2} found in the Wilsonian formulation, it arises in a very different (and much simpler) way. In particular, although the starting point looks the same, \eqref{CME2general} contains $K$ regularisation in the antibracket. 

At $O(g^3)$, we would need to solve $s_0\, \Gamma_{3,\text{cl}} = -(\Gamma_1,\Gamma_{2,\text{cl}})$ for some local classical five-point vertex $\Gamma_{3,\text{cl}}$. Since that is not possible by dimensions, it better be that $(\Gamma_1,\Gamma_{2,\text{cl}})$ already vanishes. This is straightforward to confirm using \eqref{sol1} and \eqref{sol2c}. In fact, this is nothing but the final term in \eqref{sigma30}.

We note that we have now confirmed the identification of $\Gamma_{I,\text{cl}}$ with the 1PI part of $S_{I,\text{cl}}$ as used below \eqref{treeExpansion}.

For $n\cu>3$, $\Gamma_{n,\text{cl}}$ must also vanish by dimensions, and this is consistent with 
the classical master functional
\be 
s_0\,\Gamma_n +\half \sum_{m=1}^{n-1} (\Gamma_{n-m},\Gamma_m)\,,
\ee
which now must also vanish by dimensions (\cf \ref{point3}), as is also true directly by iteration.

\subsection{One loop equations}
\label{sec:oneloopG}

To second order and one loop, the flow equation \eqref{GammaFlow} and mST \eqref{mST} now read
\beal
\label{flow2G}
\dot{\Gamma}_{2,\text{q}} &=  \half\, \text{Str}\left( -\dot{\propH} \Gamma^{(2)}_1 \propH \Gamma^{(2)}_1 + \dot{\propH}\Gamma^{(2)}_{2,\text{cl}}\right)\,, \\
\label{mST2}
s_0\, \Gamma_{2,\text{q}} &= \Delta \Gamma_{2,\text{cl}} - \text{Tr}\left( K  \Gamma^{(2)}_{1*}\propH\Gamma^{(2)}_1\right)\,.
\eeal
We recognise that these equations coincide with \eqref{S2qRelations} since the term in brackets in \eqref{flow2G} contains the second functional derivative of \eqref{S2cl}:
\be 
S^{(2)}_{2,\text{cl}} = -S^{(2)}_1 \propH S^{(2)}_1 + \Op^{(2)}_{2} +\cdots\,,
\ee
where the ellipses refer to terms where $S_1$ is differentiated three times and which vanish (by tr$T^a=0$) on contraction with the differentiated propagator $\dot{\propH}$ to form $a_1$ as defined in \eqref{flow}. Similarly the second term in \eqref{mST2} coincides with the action of $\Delta$ on \eqref{S2redDef}. The solution is therefore just a renaming of $S_{2,\text{q}}$ in \eqref{S2q}, \ie
\be 
\label{G2q}
\Gamma_{2,\text{q}} = \half\,C_A\, A_\mu\, \mathcal{A}_{\mu\nu}(\partial)\, A_\nu + C_A\,A^*_\mu\, \mathcal{B}(-\partial^2)\, \partial_\mu C\,,
\ee
where $\mathcal{A}_{\mu\nu}$ obeys \eqref{AWI}. 

Due to the cyclicity of the supertrace, \eqref{flow2G} is immediately integrable:
\be
\label{sol2}
\Gamma_{2,\text{q}} =  \text{Str}\left(-\tfrac14\propH\Gamma^{(2)}_1\propH\Gamma^{(2)}_1
+\tfrac12 \propH \Gamma^{(2)}_{2,\text{cl}}\right)\,,
\ee
which of course just comes from the expansion of the closed-form one-loop solution to \eqref{GammaFlow}:
\be 
\label{GammaOneLoop}
\Gamma_\text{q} = \half \,\text{Str} \ln \left( \propH^{-1} + \Gamma^{(2)}_{\text{cl}} \right)\,.
\ee
Expanding further, we can therefore also write down the solution to the flow equation at third order, and thus we have, together with expanding the mST \eqref{mST} to third order and one loop:
\beal
\label{sol3}
\Gamma_{3,\text{q}} &= \text{Str}\left( -\tfrac12 \propH\Gamma^{(2)}_1\propH \Gamma^{(2)}_{2,\text{cl}} +\tfrac16 
\propH\Gamma^{(2)}_1\propH\Gamma^{(2)}_1\propH\Gamma^{(2)}_1\right)\,, \\
\label{mST3}
s_0\, \Gamma_{3,\text{q}} &= -(\Gamma_1,\Gamma_{2,\text{q}}) + \text{Tr}\left( -K  \Gamma^{(2)}_{1*}\propH\Gamma^{(2)}_{2,\text{cl}}
+ K  \Gamma^{(2)}_{1*}\propH\Gamma^{(2)}_1\propH\Gamma^{(2)}_1\right)\,.
\eeal

The expressions \eqref{sol2} -- \eqref{sol3} for the one-loop 1PI effective action are IR regulated as they must be since they must have derivative expansions. They should be understood to be supplied with integration constants, which are thus also derivative expansions but independent of $\Lambda$. However, since the only scale is $\Lambda$, by dimensions all but a small number of integration constants must actually vanish. 

The non-trivial functional dependence on $\Lambda$, as specified by the body of the momentum integrals that make up the one-loop vertices, is thus unambiguous. For such momentum integrals the mST identities \eqref{mST}, \eqref{mST2} and \eqref{mST3}, must therefore already be satisfied, as we claimed in \ref{point4}. We will confirm this in sec. \ref{sec:oneloopthirdorder}.

The $\Lambda$-integration constants that do not necessarily vanish, multiply marginal and relevant operators, whose coefficient integrals 
are typically UV divergent. 
In these cases the integration constants must be chosen to cancel the divergence so that we get finite solutions as required. Nevertheless in the resulting finite solution, since by dimensions  the relevant operators necessarily have $\Lambda$-dependent coefficients, their integration constants are determined uniquely. (We will see examples in sec. \ref{sec:oneloopsols}.) Only for the marginal operators is there some arbitrariness, where the integration constants must also be adjusted so as to respect the mST identities. This is the subject of the next section.

\section{BRST structure and renormalization in the presence of the effective cutoff}
\label{sec:general}


Now we note that we are dealing with a theory that is well defined at non-exceptional momenta {(\ie external momenta such that none of their partial sums vanish or are null)}. Consider for example a two-point vertex that depends only on one external momentum $p$. The theory thus has the property that, properly renormalized,  no IR divergences appear as $\Lambda\to0$, provided that we keep the Euclidean momentum $p\ne0$. Then we see that the UV subtractions provided by the integration constants such as to render the vertices finite at $\Lambda\ne0$, can be chosen to render the vertices finite at non-exceptional momenta in the limit $\Lambda\to0$. We will confirm this in detail in sec. \ref{sec:oneloopsols}.

For the marginal operators, we will furthermore be forced by dimensions to include $\ln\!\mu$ dependence from cancelling off logarithmic UV divergences. It is this $\ln\!\mu$ dependence that introduces arbitrariness into the $\Lambda$-integration constants and that we need to keep track of by renormalization conditions.

As can be seen from  \eqref{mST2} and \eqref{mST3}, the parts that have this arbitrariness must be $s_0$--invariant. For the two-point vertices we have two solutions:
\be
\label{arbitraryTwopoint}
\half A_\mu ( -\Box \delta_{\mu\nu} +\partial_\mu\partial_\nu) A_\nu\qquad\text{and}\qquad A^*_\mu\partial_\mu C\,,
\ee
while for three-point vertices there is only one solution, namely a multiple of \eqref{sol1}. The higher-point vertices have no freedom in the solution (as we already noted in secs. \ref{sec:secondWC}, \ref{sec:thirdorderWC} and \ref{sec:secondGC}).

Let us spell out how this implies, despite the presence of a cutoff $K$,  the same key steps that one obtains in the proof of renormalizability of gauge theories using the Zinn-Justin equation, \aka CME: $(\Gamma,\Gamma) = 0$.
Altering the coefficients in front of \eqref{arbitraryTwopoint} and \eqref{sol1} induces changes in the rest of the solution. Indeed from inspection of \eqref{mST3}, we see that addition of an $s_0$-closed perturbation $\delta \Gamma_{2,\text{q}}$ requires a change $\delta\Gamma_{3,\text{q}}$ such that
\be s_0 \,\delta\Gamma_{3,\text{q}} := (\delta\Gamma_{3,\text{q}},\Gamma_0) = -(\Gamma_1,\delta\Gamma_{2,\text{q}})\,. \ee
Since the classical solutions for $\Gamma_0$ and $\Gamma_1$ are unchanged at the quantum level, the above is just the $g^3$ part of the demand that any change to the one-loop $\Lambda$-integration constants result in an operator $\delta\Gamma^{(\ell=1)} \equiv \delta\Gamma_\text{q}$ that is invariant under the
full classical BRST transformation, \eqref{fullBRSTG}, \ie
\be 
\label{Zinn1}
s_\text{cl}\, \delta\Gamma^{(\ell)} := (\delta\Gamma^{(\ell)},\Gamma_\text{cl}) = 0\,.
\ee
For $\ell=1$, this is clearly true because at one loop the rest of the mST \eqref{mST} depends only on the classical {effective} action. Summing the contributions \eqref{sol0}, \eqref{sol1}, \eqref{sol2c} (there being no more \cf sec. \ref{sec:secondGC}), the classical effective average action \eqref{Gamma} is given in total by
\be 
\label{Gammacl}
\Gamma_\text{cl} = \Gamma_\text{0}+g\Gamma_\text{1}+g^2\Gamma_\text{2,cl} = -ig C^*C^2 +A^*_\mu D_\mu C+
\tfrac14 F^2_{\mu\nu} \,,
\ee
\ie just the usual expression for the classical action complete with antifield sources for the classical BRST transformations\footnote{$Q_\text{cl}$ preserves antighost number, but this is somewhat accidental. To exploit antighost number with interacting charges generally requires regrading since antighost number is not conserved. \textit{E.g.} from \eqref{fullBRSTG}, $Q^-_\text{cl} C^*$ now has a part that does not lower antighost number, while from \eqref{S2red}, we see that  $Q_{2,\text{cl}} a_\mu$ has a part that raises it by one.}
$Q_\text{cl}C$ and $Q_\text{cl}A_\mu$, \cf \eqref{fullBRSTG},
where the covariant derivative and field strength are defined as:
\be 
D_\mu C = \partial_\mu C -ig [A_\mu,C]\qquad \text{and}\qquad F_{\mu\nu} = \tfrac{i}g [D_\mu,D_\nu]\,.
\ee

In general, $\delta\Gamma^{(\ell)}$ contains all the (changes in) $\Lambda$-integration constants at the $\ell$ loop level. As we already noted in sec. \ref{sec:five}, these integration constants play the r\^ole of counterterms in the standard treatment. Indeed it is clearly true that at $\ell$ loops, they must also satisfy \eqref{Zinn1}, just as one would deduce from the Zinn-Justin equation \cite{ZinnJustin:1974mc,ZinnJustin:1975wb,ZinnJustin:2002ru}, since the rest of the mST  depends only on lower loop orders, either through
$ 
\frac12 \sum^{\ell-1}_{\jmath=1} (\Gamma^{(\ell-\jmath)},\Gamma^{(\jmath)})
$,
or because the correction term in \eqref{mST} contributes one extra loop (equivalently one extra factor of $\hbar$).

Although we have derived this for $\Gamma$, closely similar arguments apply to the Wilsonian action $S$, as we have already intimated in sec. \ref{sec:five}. Indeed the part $\delta S^{(\ell)}$ containing the $\Lambda$-integration constants must, by the QME \eqref{QMF}, satisfy the analogous equation at $\ell$-loop order:
\be 
s_\text{cl}\, \delta S^{(\ell)} := (\delta S^{(\ell)}, S_\text{cl}) = 0\,,
\ee
since the measure operator $\Delta$ also supplies an extra loop. Furthermore, \eqref{CMEequivalence} implies that any violations of $\Sigma=0$ would be equal under the Legendre transform relation:
\be 
s_\text{cl}\, \delta S^{(\ell)} := (\delta S^{(\ell)}, S_\text{cl}) = (\delta\Gamma^{(\ell)},\Gamma_\text{cl}) =: s_\text{cl}\, \delta\Gamma^{(\ell)} \,.
\ee
A major difference however is that $S_\text{cl}$ is not a closed expression, but rather contains contributions to all orders in $g$ (this being consistent with the fact that its antibracket \eqref{QMEbitsReg} carries $K$ regularisation).

To bring the kinetic terms back to {normaliz}ed form, as defined by some renormalization condition, we need to apply wavefunction renormalization, \ie a rescaling of the (anti)fields. Order by order in the coupling, or loop expansion, the rescaling will then provide us with the corresponding choices for the $\Lambda$-integration constants. 
We have seen that these perturbative contributions must be closed under the classical BRST transformations. This implies that the wavefunction rescaling must in fact be a (classical) canonical transformation, and thus of the particularly simple form:
\be 
\label{canonicalK}
\mathcal{K} = Z^{\half}_{E}\Phi^*_E\Phi^E_{(r)}\,,
\ee
where summation over the repeated indices should still be understood as determined by the (anti)fields while $Z$ however also depends on (anti)field flavour, and where we use the general recipe for a finite classical canonical transformation (see \eg \cite{Gomis:1994he}):
\be 
\label{canonical}
\Phi^E = \frac{\partial_l}{\partial\Phi^*_E} \mathcal{K}[\Phi_{(r)},\Phi^*]\,,\qquad \Phi^*_{(r) E} = \frac{\partial_r}{\partial \Phi^E_{(r)}} \mathcal{K}[\Phi_{(r)},\Phi^*]\,,
\ee
the subscript $(r)$ labelling the renormalized (anti)fields.
Thus the fields and antifields renormalize in opposite directions: 
\be 
\label{wavefunctionGI}
A_\mu = Z^{\half}_A A_{(r)\mu}\,,\quad A^*_\mu = Z^{-\half}_A A^*_{(r)\mu}\,,\quad C = Z^{\half}_C C_{(r)}
\,,\quad C^* = Z^{-\half}_C C^*_{(r)}\,.
\ee
The freedom in the $\Lambda$-independent part of the solution has therefore been parametrised as
\besp \label{renormParam}
\frac12 Z^{-1}_A  A_\mu ( -\Box \delta_{\mu\nu} +\partial_\mu\partial_\nu) A_\nu\ +\ Z^{\half}_AZ^{-\half}_C A^*_\mu\partial_\mu C 
\ -\ i g Z^{-1}_g Z^{-\half}_C \left( C^*C^2+A^*_\mu[A_\mu,C]\right) \\ 
-igZ^{-1}_gZ^{-\tfrac32}_A \partial_\mu A_\nu [A_\mu,A_\nu]
\ -\ \frac14 g^2 Z^{-2}_g Z^{-2}_A [A_\mu,A_\nu]^2\,.
\eesp
Here we have also introduced $Z_g$ for renormalizing the coupling, which thus takes care of the separate freedom that appears at the three-point level. {The above equation appears with the inverse transformation so that on applying \eqref{wavefunctionGI} and $g = Z_g \,g_{(r)}$ the result appears in {normaliz}ed form in terms of the renormalized fields and coupling.}

We note that at one loop we will find that all these factors take the form 
\be 
\label{Zn}
Z_n = 1+ g^2 z_n\,, 
\ee 
where $n=A,C,g$, and in fact the 
\be 
\label{zngeneral}
z_n = \gamma_n \ln(\mu/\Lambda) + z^0_n\,,
\ee
where the $\gamma_n$ are computable (as we will see) and the freedom is here overparametrised in the choice of $\mu$ and the $z^0_n$. 
We also note that it is straightforward to verify that perturbative contributions to \eqref{renormParam} do satisfy \eqref{Zinn1}. For example from \eqref{Zn}, the one-loop  $O(g^2)$ part of \eqref{renormParam} parametrises the change in the two-point vertices (kinetic terms) as $\delta \Gamma_{\q,2} =  s_0\, ( g^2\mathcal{K}_2)$, where 
\be 
\mathcal{K}_2 = \half\, z_A\, A^*_\mu A_\mu +\half\, z_C\, C^* C\,.
\ee

Again, closely similar arguments apply to the Wilsonian action. Apart from the freedom $Z_g$ to adjust the coefficient of $S_1$, we only have the freedom to alter the analogous kinetic terms to \eqref{arbitraryTwopoint}.
Thus these latter $\Lambda$-integration constants must also be parametrised by a canonical transformation:
\be 
\label{canonicalS}
\Ph^E = \frac{\partial_l}{\partial\Ph^*_E} \tilde{\mathcal{K}}[\Ph_{(r)},\Ph^*]\,,\qquad \Ph^*_{(r) E} = \frac{\partial_r}{\partial \Ph^E_{(r)}} \tilde{\mathcal{K}}[\Ph_{(r)},\Ph^*]\,,
\ee
(the antibracket is invariant under this by statistics and thus also with the regularisation \eqref{QMEbitsReg} \cite{\morris}),
where now 
\be 
\label{canonicalKS}
\tilde{\mathcal{K}} = Z^{\half}_{E}\Ph^*_E\,\Ph^E_{(r)}\,,
\ee
and thus 
\be 
\label{wavefunctionGIS}
a_\mu = Z^{\half}_A a_{(r)\mu}\,,\quad a^*_\mu = Z^{-\half}_A a^*_{(r)\mu}\,,\quad c = Z^{\half}_C \,c_{(r)}
\,,\quad c^* = Z^{-\half}_C c^*_{(r)}\,.
\ee
Note that these $Z$ factors are indeed the same as the ones in the 1PI effective action, as might be expected given that $\Gamma_I$ provides the 1PI part of $S$. See app. \ref{app:Zs} for a proof.


The changes of variables, \eqref{wavefunctionGI} and \eqref{wavefunctionGIS}, do not leave the cutoffs and thus neither the flow equations,  \eqref{GammaFlow} and \eqref{flow}, nor the Legendre transform identity \eqref{Legendre}, invariant.
However as we have been emphasising (and will see particularly in the remainder of the paper), {the flow equations} can be solved for directly in terms of finite, and thus already renormalized, quantities. What this lack of invariance means however is that we can only enforce renormalization conditions at some fixed  (finite) scale, for example at $\Lambda=\mu$. The simplest choice then is to set 
\be 
\label{znCondition}
Z_n =1\qquad\text{at}\qquad\Lambda=\mu\,,
\ee
so that in \eqref{zngeneral} we have just
\be 
\label{zn}
z_n = \gamma_n \ln(\mu/\Lambda)\,.
\ee
{For finite values of $\Lambda\ne\mu$, the effective action is still finite but the kinetic terms appear as in \eqref{renormParam} and are not in normalized form. Eqns. \eqref{wavefunctionGI} then provide us with the further finite renormalization that would be required to bring the kinetic terms back to normalized form.}

Let us acknowledge that one can start instead with altered flow equations, both for the Wilson/Polchinski effective action and for the 1PI effective action, such that they depend on running anomalous dimensions $\gamma_n(\Lambda)$ that are determined by setting  renormalization conditions that hold at all values of $\Lambda$. However in this case the map between the two effective actions is no longer as simple as \eqref{Legendre} \cite{Rosten:2010pc,Rosten:2011mf}.

For completeness we also consider gauge invariant extended basis and gauge fixed basis. Although we display only the 1PI formulation, again the analogous equations hold for the Wilsonian action.
From sec. \ref{sec:S0}, we see that in gauge invariant extended basis we also include in $\Gamma$,
\be 
\label{extended}
\frac{\xi}{2} B^2 + \bar{C}^* B\,,
\ee
however these terms are not generated by quantum corrections so do not need renormalization. In gauge fixed basis we make the analogous canonical transformation,
$
\Phi^*_A |_\text{gf} = \Phi^*_A |_\text{gi} + \partial^r_A \Psi
$, to \eqref{canontogf},
where the gauge fermion is the classical field equivalent of \eqref{Psi}, namely
$
\Psi = - i \,\bar{C} \partial\!\cdot\! A\,,
$
implying the transformations \eqref{GFGI}. Recall that the dependence on $\bar{C}$ is only through the combination $A^*_\mu - i \partial_\mu\bar{C}$. In order that this is also renormalized, we therefore need $\bar{C}$ to renormalize like $A^*_\mu$. By \eqref{canonicalK}, this implies that $\bar{C}^*$ must renormalize like $A_\mu$. We see that these choices then also leave the gauge fermion invariant and preserve the form the transformations \eqref{GFGI}. Since \eqref{extended} receives no corrections, this determines also the renormalization of $B$ and $\xi$. In summary, we must supplement \eqref{wavefunctionGI} with:
\be 
\label{wavefunctionGF}
\bar{C} = Z^{-\half}_A \bar{C}_{(r)}\,,\quad \bar{C}^* = Z^{\half}_A \bar{C}^*_{(r)}\,,\quad
B = Z^{-\half}_A B_{(r)}\,,\quad \xi = Z_A\, \xi_{(r)}\,.
\ee
Since integrating out $B$ gives the gauge fixing term $(\partial\cu\cdot A)^2/\xi$, we see that the $\xi$ renormalization is consistent with standard methods where one finds that the gauge-fixing term does not renormalize.

These wavefunction renormalization factors \eqref{wavefunctionGI} and \eqref{wavefunctionGF} (similarly \eqref{wavefunctionGIS} \etc) are not the standard ones,  because   \eqref{renormParam} already solves the Slavnov-Taylor wavefunction renormalization identities. Although $Z_A$ is $Z_3$ in standard notation, both $Z_C$ and $Z_A$ re{normaliz}e the (anti)ghost fields.  In a standard parametrisation (\eg \cite{Itzykson:1980rh}) we would identify 
\be 
\label{standard}
Z_3 = Z_A\,,\quad \tilde{Z}_3 = Z_C^{\half} Z_A^{-\half}\,,\quad Z_1 = Z_g Z^{\tfrac32}_A\,,\quad \tilde{Z}_1= Z_g Z_C^{\half}\,,\quad Z_4 = Z^2_gZ^2_A\,,
\ee
where $\tilde{Z}_3$ is the wavefunction renormalization factor for both $C$ and $\bar{C}$, and $Z_1$, $\tilde{Z}_1$ and $Z_4$ are the factors for the $A^3$, $\bar{C} A C$ and $A^4$ vertices respectively. Then the following fractions are all equal to $Z_g Z^{\half}_A$:
\be 
\label{SlavnovTaylor}
\frac{Z_1}{Z_3} = \frac{\tilde{Z}_1}{\tilde{Z}_3} = \frac{Z_4}{Z_1}\,.
\ee
These are indeed the Slavnov-Taylor identities guaranteeing the universality of the gauge coupling. Thus we have verified \ref{point5}.

\section{One loop 1PI solutions}
\label{sec:oneloopsols}

In the last section, we explained how  a restricted set of wavefunction renormalization factors $Z_n$ appear naturally in the finite continuum solution of the flow equations in such a way as to satisfy the QME \eqref{QMF} in the presence of a cutoff, or equivalently the mST \eqref{mST}, and thus satisfy the corresponding Slavnov-Taylor identities. In sec. \ref{sec:oneloopG} we used the compact form of the 1PI effective action (\aka IR-cutoff Legendre effective action or effective average action), to list up to $O(g^3)$, expressions for the 1PI vertices and the mST identities they must satisfy. In this section we now compute these vertices as momentum integrals, show how the requirement of a derivative expansion ensures a smooth limit to the standard expressions in the limit in which the IR cutoff $\Lambda\to0$, and extract the $Z_n$ factors. We use them to verify the standard form of the one-loop Yang-Mills $\beta$ function, here computed as the flow with respect to $\Lambda$.

\subsection{Second order in coupling and wavefunction renormalization}
\label{sec:oneloopsecondorder}

From \eqref{G2q} and \eqref{sol2} one obtains\footnote{We find that the calculations here and later are facilitated by using the vertices constructed in app. \ref{app:vertices}.}
\be
\label{A}
\mathcal{A}_{\mu\nu}(p) = \int_{q} \Big\{ \propH(q)\propH(q\cu+p)\,q_\mu (q\cu+p)_\nu -\half\,\propH_{\alpha\beta}(q)\propH_{\rho\sigma}(q\cu+p)\,\Theta_{\rho\alpha\mu}\Theta_{\sigma\beta\nu} +\propH_{\alpha\alpha}(q)\,\delta_{\mu\nu} -\propH_{\mu\nu}(q)
\Big\}\,,
\ee
where, using \eqref{sol1}, \eqref{GIbasisDeriv} and \eqref{propagators}, the first two terms come from the ghost propagator and gauge field propagator contributions to the first term in \eqref{sol2}, and from \eqref{Theta} we have written 
\be 
\Theta_{\rho\alpha\mu} = (p \cu- q)_{\sigma} \delta_{\beta\nu}
- (2p\cu+q)_{\beta}\delta_{\nu\sigma} + \delta_{\beta\sigma}(p\cu + 2q)_{\nu}\,.
\ee
The last two terms in \eqref{A} come from the second term in \eqref{sol2}.  As explained at the end of sec. \ref{sec:oneloopG}, we should understand  \eqref{A} as supplied with $\Lambda$-integration constants which in particular ensure that the result is UV finite. 
Thanks to the IR regularisation, \cf also \eqref{propmunu}, the propagators have a Taylor expansion in $p^\mu$ for any $q$ and thus we confirm that \eqref{A} also has a Taylor expansion in $p^\mu$. Decomposing $\mathcal{A}_{\mu\nu}(p)$ into its longitudinal and transverse parts,
\be 
\label{Asplit}
\mathcal{A}_{\mu\nu}(p) = \mathcal{A}^L(p^2)\, P^L_{\mu\nu} + \mathcal{A}^T(p^2)\, P^T_{\mu\nu}\,,\quad\text{where}\quad P^L_{\mu\nu} = p_\mu p_\nu/p^2\ \text{and}\  P^T_{\mu\nu} = \delta_{\mu\nu} - P^L_{\mu\nu}\,,
\ee
by dimensions and the Taylor expansion property, we have that
\be 
\label{ATaylor}
\mathcal{A}^I(p^2) =\Lambda^2 \sum_{n=0}^\infty \alpha^I_n  \left(\frac{p^{2} }{\Lambda^{2}}\right)^n\,,\qquad I=L,T\,,
\ee
where the coefficients $\alpha^I_n=\alpha^I_n(\xi)$ are dimensionless and well defined, and $\alpha^L_0=\alpha^T_0 = \alpha_0$ are equal. In particular this zeroth order provides a $\Lambda$-dependent `mass' term in \eqref{G2q}:
\be 
\label{AL0}
C_A\, \mathcal{A}_{\mu\nu}(0) = \alpha_0\, C_A\, \Lambda^2 \delta_{\mu\nu}\,.
\ee
As advertised at the end of sec. \ref{sec:oneloopG}, we see that except for the marginal $\alpha_1^I$ coefficients, the $\Lambda$-integration constants are already determined uniquely as they must be by dimensions. Indeed from \eqref{A}, the irrelevant pieces ($a^I_{n>1}$) are given by well defined momentum integrals since these integrals are also UV finite, while $\alpha_0$ can be defined by subtracting from $\mathcal{A}_{\mu\nu}(0)$ as defined in \eqref{A}, the same expression evaluated at $\Lambda=0$. On the other hand, the form of the marginal transverse part has been fixed by the renormalization condition \eqref{znCondition} to
\be 
\label{gammaAdef}
C_A\, \alpha^T_1 = -\gamma_A \ln(\mu/\Lambda)\,, 
\ee
as follows from \eqref{renormParam}, \eqref{Zn} and \eqref{zn}.

Concentrating now on the longitudinal part, we use the symmetry of the first two terms in \eqref{A}  under $q\mapsto-q\cu-p$. (For more details see app. \ref{app:XYZ}.)
Eliminating propagators when the same factor appears in the numerator, including eliminating the dot product in
$p\cu\cdot q/(p\!+\!q)^2$ by expressing $p\!\cdot\!q =\half (p\!+\!q)^2 -\half (p^2+q^2)$, and converting terms with single propagators of form $1/(p\cu+q)^2$ to ones with $1/q^2$ by using $q\mapsto-q\cu-p$, the end result 
is thus uniquely expressed:
\be 
\label{ALplus}
\mathcal{A}^L(p^2) = \frac34\, (\xi+3)\! \int_q\! \frac{\bar{K}(q)}{q^2}
\ -\  \int_q\! \frac{\bar{K}(q)\,\bar{K}(p\cu+q)}{q^2}\left( (1-\xi) \frac{(p\!\cdot\!q)^2}{p^2q^2} + 4\frac{p\!\cdot\!q}{p^2}+\xi+2\right)\,. 
\ee
This expression is still IR regulated, but
at first sight it appears to include unregulated UV divergences. However the $K$-independent part vanishes (this and similar later manipulations need some care, see app. \ref{app:Care}):
\be 
\label{vanish}
\left\{ \frac34\, (\xi+3) -\frac14(1-\xi)-\xi-2\right\} \int_q \frac1{q^2} = 0\,.
\ee
This means \eqref{ALplus} is equal to 
\be 
\label{ALplusplus}
\mathcal{A}^L(p^2) = -\frac34\, (\xi+3)\! \int_q\! \frac{{K}(q)}{q^2}
\ +\  \int_q\! \left(\propH(q) K(p\!+\!q)+\frac{K(q)}{q^2}\right) \left( (1-\xi) \frac{(p\!\cdot\!q)^2}{p^2q^2} + 4\frac{p\!\cdot\!q}{p^2}+\xi+2\right)\,.
\ee
Collecting the $K(q)/q^2$ pieces, they cancel exactly as in \eqref{vanish}, and thus we are left only with an expression identical to \eqref{F}. Since the modified {Ward-Takahashi} identity \eqref{AWI} tells us $\mathcal{A}^L = \mathcal{F}$, we see that it holds identically as a statement about momentum integrals, confirming point \ref{point4} in this example.

Since \eqref{F} is both IR and UV regulated, $\mathcal{A}^L$ has now been cast in a form which has a  Taylor expansion in $p^2$ with well defined coefficients. As examples, we find for the mass term \eqref{AL0}, and the marginal $p_\mu p_\nu$ term in $\mathcal{A}_{\mu\nu}(p)$, \ie $O(p^2)$ longitudinal part:
\beal 
\alpha_0 &= \frac{1}{4(4\pi)^2}\int^\infty_0\!\!\!\!du\,\left\{ (1+3\xi)\,K(u)-(5+3\xi)\,K^2(u)\right\}\,,\nn\\
\alpha^L_1 &= \frac1{(4\pi)^2}\left\{ -\frac16-\frac14\xi+\frac{5+\xi}{4}\int^\infty_0\!\!\!\!du\, u\left[K'\!(u)\right]^2\right\}\,,
\label{mass}
\eeal
where prime is differentiation with respect to $u$, and we recall that $K(q)$ is really $K(u=q^2/\Lambda^2)$. We note that the $K$-dependent pieces give integrals that are finite but non-universal. 

We also note that, due to the finiteness of \eqref{F}, $\alpha^L_1$ has no logarithmic running, consistent with the general arguments that led to \eqref{renormParam}. In principle, having shown that the momentum integral in 
\eqref{ALplusplus} is equal to $\mathcal{F}(p^2)$, we might expect to write $\mathcal{A}^L(p^2) = \mathcal{F}(p^2)+\mathcal{A}^L_0$, where $\mathcal{A}^L_0$ is the $\Lambda$-integration constant. Then the modified {Ward-Takahashi} identity \eqref{AWI} would just tell us that $\mathcal{A}^L_0$ vanishes. However since \ref{point4} tells us that the identity holds as a statement about momentum integrals, and the momentum integrals in this case can be cast in a form in which they are well defined, there is clearly no reason to introduce such a $\Lambda$-integration constant in this case.

Finally we note that from \eqref{F}, if we stay at $p\ne0$ we can safely take the limit $\Lambda\to0$, thus defining the longitudinal part of the one-loop  two-point vertex of the standard physical Legendre effective action. For non-zero $p$ however, the $K(p\!+\!q)$ factor in \eqref{F} forces the result to vanish in this limit. Thus we find that the physical one-loop two-point vertex is purely transverse, in agreement with standard results.

To evaluate $\mathcal{A}^T$, and $\mathcal{B}$ in \eqref{G2q}, it is helpful to provide UV regularisation.  We will demonstrate this for $\mathcal{B}$. For $\mathcal{A}^T$ we work instead with its RG time derivative, which we could also directly extract from \eqref{flow2G}. This is well defined in both the UV and the IR since $\dot{\bar{K}}=-\dot{K}$ provides UV regularisation. The $\dot{\mathcal{A}}^T(0)$ piece confirms the mass term $\alpha_0$ in \eqref{mass} while  the $O(p^2)$ part yields
\be 
\left(2\xi-\frac{26}3\right) \int^\infty_0\!\!\!\!\!\!du\, \bar{K}'\!(u)\, \bar{K}(u)
\ee
(up to vanishing surface terms that depend on momentum routing) and thus from \eqref{gammaAdef} and the normalization conditions below \eqref{propIR}, we get
\be 
\label{ga}
\gamma_A = \left( \frac{13}{3}-\xi\right) \frac{C_A}{(4\pi)^2}\,.
\ee

For $\mathcal{B}$, using again \eqref{sol1}, \eqref{GIbasisDeriv} and \eqref{propagators}, we get from \eqref{G2q} and \eqref{sol2}:
\be 
p^{2} {\cal B}(p^2) = - 
\int_{q} \propH_{\mu\nu}(q) \propH(q\cu+p) p_{\mu}(q\cu+p)_{\nu}
=\int_{q}\propH(q)\propH(q\cu+p) \Bigl[
  p^{2} + \xi(p\cdot q) + (\xi-1) \frac{(p\cdot q)^{2}}{q^{2}} \Bigr]\,.
  \label{cal_B3}
\ee
We note that $\mathcal{B}(p^2)$ has a derivative (\aka Taylor) expansion. However $\mathcal{B}(0)$  has a logarithmic UV divergence, which will thus be cancelled by the dimensionless $\Lambda$-integration constant $\mathcal{B}_0$. Making this explicit, and converting the integrand in the same way as above \eqref{ALplus}, we get:
\be 
\label{Bintegral}
\mathcal{B}(p^2) =  \int_q\! \frac{\bar{K}(q)\,\bar{K}(p\cu+q)}{q^2}\left( -\frac1{2(p\cu+q)^2} +\frac{1-\xi}4\frac{p^2}{q^2(p\cu+q)^2}+\frac{1-\xi}2\frac{q\cu\cdot p}{p^2q^2} +\frac{\xi-1}{4\,q^2}\right)+\mathcal{B}_0\,.
\ee
Thus we find that (see app. \ref{app:Care}): 
\be 
\label{Batz}
\mathcal{B}(0) = \frac{\xi-3}{4}\int_q\! \frac{\bar{K}^2(q)}{q^4}\ +\mathcal{B}_0\ =\ \frac{\gamma_A-\gamma_C}{2C_A}\ln\frac{\mu}{\Lambda}\,,
\ee
where the last equality follows from \eqref{renormParam} and the renormalization conditions \eqref{Zn} and \eqref{zn}. From the RG time derivative it is almost immediate to compute $\gamma_A-\gamma_C$:
\be 
\dot{\mathcal{B}}(0) = \frac{3-\xi}{2}\int_q \frac{\bar{K}(q)\dot{K}(q)}{q^4} = \frac{\xi-3}{(4\pi)^2} \int^\infty_0\!\!\!\!du\, \bar{K}(u) \bar{K}'(u) = \frac{\xi-3}{2(4\pi)^2}\,,
\ee
(evidently the result is effectively the same as in dimensional regularisation) and thus, 
\be 
\label{gagc}
\gamma_A -\gamma_C = \frac{\xi-3}{(4\pi)^2}\,C_A\,.
\ee
Keeping $p\ne0$ and letting $\Lambda\to0$, we get the physical vertex. We see that the momentum integral in \eqref{Bintegral} then coincides with 
its $K$-independent part:\footnote{Here we recognise that the third term in brackets in \eqref{Bintegral} vanishes, and that the second and fourth terms can be combined to eliminate their respective IR divergences.}
\be 
\label{Bphys}
\mathcal{B}_\text{phys}(p^2) = \int_q\!\left( \frac{\xi-3}{4}\frac{1}{q^2(q\cu+p)^2}+\frac{\xi-1}{2}\frac{q\cu\cdot p}{q^4(q\cu+p)^2}\right)+\mathcal{B}_0\,.
\ee
To evaluate this completely it is helpful 
to provide some UV regularisation. It does not matter what regularisation we choose. Different regularisations will give a different finite part to the integral which however then implies a different $\mathcal{B}_0$ such that \eqref{Batz} remains satisfied. We use dimensional regularisation, setting spacetime dimension $d=4-2\epsilon$ with $\epsilon>0$. Computing the integral in \eqref{Batz}:
\be 
\int_q \frac{\bar{K}^2(q)}{q^4} = \frac1{\Gamma(2-\epsilon) \Lambda^{2\epsilon}(4\pi)^{2-\epsilon}}\int^\infty_0 \!\!\!\!du\, \frac{\bar{K}^2(u)}{u^{1+\epsilon}} \,.
\ee
Splitting the last integral at $u=1$ we have
\be
\int^\infty_0 \!\!\!\!du\, \frac{\bar{K}^2(u)}{u^{1+\epsilon}}  = \int^1_0 \!\!\!\!du\, \frac{\bar{K}^2(u)}{u^{1+\epsilon}}\ +\ \int^\infty_1 \!\!\!\!du\, \frac{\bar{K}^2(u)-1}{u^{1+\epsilon}}\ + \ \int^\infty_1 \!\!\frac{du\phantom{1}}{u^{1+\epsilon}}\,.
\ee
In the first and second integrals on the RHS, the $\epsilon\to0$ limit can now be safely taken. The final integral can be done exactly. Substituting into \eqref{Batz} we thus find
\be 
\label{Bzero}
\mathcal{B}_0 = \frac{3-\xi}{4(4\pi)^2}C_A\left\{ \frac{1}{\epsilon}+1+\ln(4\pi)-\gamma_E-\ln(\mu^2)+\int^1_0\!\frac{du}{u}\bar{K}^2(u) + \int^\infty_1\!\frac{du}{u}\left( K^2(u) -2 K(u)\right)
\right\}\,,
\ee
where $\gamma_E$ is Euler's constant, and terms that vanish as $\epsilon\to0$ are discarded.

Computing the momentum integrals in \eqref{Bphys} by dimensional regularisation using standard methods, we confirm that \eqref{Bzero} cancels the UV divergence, and thus arrive at the final expression for the physical vertex:
\be 
\label{BphysExplicit}
\mathcal{B}_\text{phys}(p^2) =  \frac{3-\xi}{4(4\pi)^2}\, C_A \left\{ \ln\frac{p^2}{\mu^2} -1 +\int^1_0\!\frac{du}{u}\bar{K}^2(u) + \int^\infty_1\!\frac{du}{u}\left( K^2(u) -2 K(u)\right)
\right\} +\frac{1-\xi}{2(4\pi)^2}\,C_A\,.
\ee
We note that all but the $\ln(p^2/\mu^2)$ part amounts to a finite (non-universal) wavefunction renormalization. It is the price to pay for the simple expression \eqref{Batz}, following from the renormalization condition \eqref{znCondition}. In particular it is through such a renormalization condition that $\mathcal{B}_0$ becomes $K$-dependent, despite the fact that the momentum integral defining the physical vertex in \eqref{Bphys} is $K$-independent. We could have chosen a renormalization condition such that $\mathcal{B}_\text{phys}$ would be given only by the $\ln(p^2/\mu^2)$ part or such that it agrees with the $\overline{\text{MS}}$ result:
\be 
\label{BMSbar}
\mathcal{B}^{\overline{\text{MS}}}_\text{phys}(p^2) =  \frac{C_A}{(4\pi)^2} \left\{\frac{3-\xi}4 \ln\frac{p^2}{\mu^2} -1 \right\}\,,
\ee
for examples. Then the $z^0_n$  in \eqref{zngeneral} would have had finite corrections.

Let us note that, when expressed in standard form \eqref{standard}, the equations \eqref{ga} and \eqref{gagc} give the same anomalous dimensions $\gamma_3$ and $\tilde{\gamma}_3$ that one obtains by standard techniques. Indeed this is guaranteed since the $\Lambda$ dependence determines the $\mu$ dependence in \eqref{zngeneral} by dimensions, and the $\mu$ dependence in turn determines the $p$ dependence by dimensions as \eg in \eqref{BMSbar}. This $p$ dependence better come out the same since it is physical (being related for example to the splitting functions by unitarity).

\subsection{Third order in coupling, the beta function and mST}
\label{sec:oneloopthirdorder}

\begin{figure}
\centering
\begin{fmffile}{AAA4}
\begin{fmfgraph*}(30,15)
\fmftop{t}
\fmflabel{$A$}{t}
\fmfleft{l,ul}
\fmflabel{$A$}{l}
\fmfright{r,ur}
\fmflabel{$A$}{r}
\fmf{gluon}{t,vt}
\fmf{gluon}{l,vb,r}
\fmf{gluon,right=0.5,tension=0.2}{vt,vb}
\fmf{gluon,tension=0.2,right=0.5}{vb,vt}
\end{fmfgraph*}
\end{fmffile}
\quad 
\begin{fmffile}{AAA}
\begin{fmfgraph*}(30,15)
\fmftop{t}
\fmflabel{$A$}{t}
\fmfleft{l,ul}
\fmflabel{$A$}{l}
\fmfright{r,ur}
\fmflabel{$A$}{r}
\fmf{gluon}{t,vt}
\fmf{gluon,left=0,tension=0.2}{vr,vt}
\fmf{gluon}{r,vr}
\fmf{gluon,right=0,tension=0.2}{vt,vl}
\fmf{gluon}{vl,l}
\fmf{gluon,tension=0,right=0}{vl,vr}
\end{fmfgraph*}
\end{fmffile}
\quad 
\begin{fmffile}{AAAg}
\begin{fmfgraph*}(30,15)
\fmftop{t}
\fmflabel{$A$}{t}
\fmfleft{l,ul}
\fmflabel{$A$}{l}
\fmfright{r,ur}
\fmflabel{$A$}{r}
\fmf{gluon}{t,vt}
\fmf{gluon}{vr,r}
\fmf{gluon}{l,vl}
\fmf{ghost,tension=0.15}{vt,vr,vl,vt}
\end{fmfgraph*}
\end{fmffile}
\quad \hskip30mm \vskip10mm
\begin{fmffile}{ACC1}
\begin{fmfgraph*}(30,15)
\fmftop{t}
\fmflabel{$C$}{t}
\fmfleft{l,ul}
\fmflabel{$A^*$}{l}
\fmfright{r,ur}
\fmflabel{$A$}{r}
\fmf{ghost,tension=0.5}{vt,t}
\fmf{gluon,tension=0.5}{vr,r}
\fmf{ghost,tension=0.5}{l,vl}
\fmf{ghost,tension=0.2}{vl,vt}
\fmf{gluon,tension=0}{vl,vr}
\fmf{gluon,tension=0.2}{vr,vt}
\end{fmfgraph*}
\end{fmffile}
\quad 
\begin{fmffile}{ACC2}
\begin{fmfgraph*}(30,15)
\fmftop{t}
\fmflabel{$C$}{t}
\fmfleft{l,ul}
\fmflabel{$A^*$}{l}
\fmfright{r,ur}
\fmflabel{$A$}{r}
\fmf{ghost,tension=0.5}{vt,t}
\fmf{gluon,tension=0.3}{vr,r}
\fmf{ghost,tension=0.5}{l,vl}
\fmf{gluon,tension=0.1}{vl,vt}
\fmf{ghost,tension=0}{vl,vr}
\fmf{ghost,tension=0.1}{vr,vt}
\end{fmfgraph*}
\end{fmffile}
\quad 
\begin{fmffile}{CsCC}
\begin{fmfgraph*}(30,15)
\fmftop{t}
\fmflabel{$C^*$}{t}
\fmfleft{l,ul}
\fmflabel{$C$}{l}
\fmfright{r,ur}
\fmflabel{$C$}{r}
\fmf{dots}{t,vt}
\fmf{ghost,tension=0.2}{vt,vr}
\fmf{ghost}{vr,r}
\fmf{ghost,tension=0.2}{vt,vl}
\fmf{ghost}{vl,l}
\fmf{gluon,tension=0,right=0}{vl,vr}
\end{fmfgraph*}
\end{fmffile}
\caption{One-loop vertices in $\Gamma_3$. In gauge invariant basis, $A^*$ also plays the r\^ole of the antighost.}
\label{fig:Gamma3}
\end{figure}
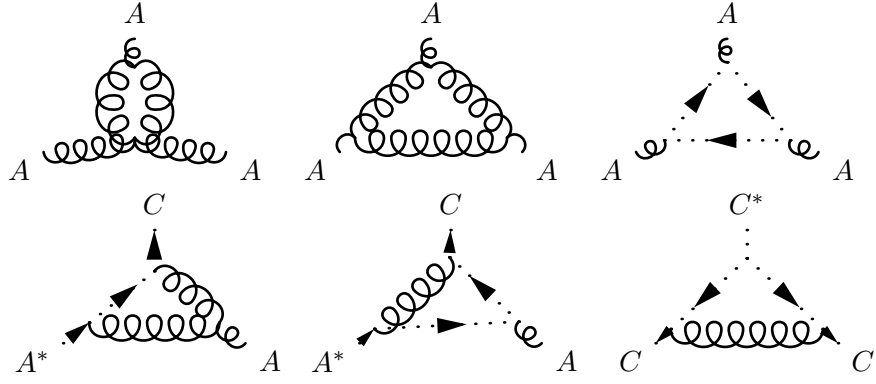

The solution \eqref{sol3} corresponds to the Feynman diagrams shown in fig. \ref{fig:Gamma3}. One can see by inspection that they are all at worst logarithmically UV divergent, and that each UV divergence is proportional to one of the three-point vertices in \eqref{renormParam}.  As discussed in sec. \ref{sec:general}, these divergences and the corresponding freedom in the choice of integration constants are constrained to satisfy the parametrisation in \eqref{renormParam}. 

\unitlength=1mm

The one-loop vertices as a whole have to satisfy the mST identities \eqref{mST3}. Apart from the freedom to choose the integration constants according to \eqref{renormParam}, there is no further flexibility, and thus the body of such a solution \eqref{sol3}, as defined by the momentum integrals, must already satisfy these mST identities as we noted in \ref{point4} and at the end of sec. \ref{sec:oneloopG}, and will confirm shortly.  We can write the vertices as
\besp 
\label{Gamma3}
\Gamma_3 = -i \int_{p,q,r} A_\mu(p) [A_\nu(q),A_\lambda(r)]\, \Gamma^{AAA}_{\mu\nu\lambda}(p,q,r)
-i \int_{p,q,r}A^*_\mu(p) [A_\nu(q),C(r)] \,\Gamma^{A^*\!\!AC}_{\mu\nu}(p,q,r)\\ -i \int_{p,q,r} C^*(p)C(q)C(r) \,\Gamma^{C^*\!CC}(p,q,r)\,,
\eesp
where $(2\pi)^4\delta(p\cu+q\cu+r)$ is understood to be included in the measure. The correction terms in the mST can similarly be written:
\besp
\label{correction}
\text{Tr}\left( -K  \Gamma^{(2)}_{1*}\propH\Gamma^{(2)}_{2,\text{cl}}
+ K  \Gamma^{(2)}_{1*}\propH\Gamma^{(2)}_1\propH\Gamma^{(2)}_1\right) = \\
\int_{p,q,r} C(p) [A_\mu(q),A_\nu(r)]\, \Delta^{CAA}_{\ \ \mu\nu}(p,q,r) + \int_{p,q,r} A^*_\mu(p) C(q) C(r) \,\Delta^{A^*\!CC}_\mu(p,q,r)\,.
\eesp 
We compute that
\be 
-(\Gamma_1,\Gamma_2) = -i C_A\left\{ \left( \{A^*_\mu,C\}+[\partial_\mu A_\nu\cu-\partial_\nu A_\mu,A_\nu]\right) \mathcal{B}\,\partial_\mu C +\partial_\mu A^*_\mu\, \mathcal{B}\, C^2 - A_\mu\, \mathcal{A}_{\mu\nu} [A_\nu,C]\right\}\,.
\ee
and thus for example from the $A^*CC$ part of \eqref{mST3} we find one must have:
\be 
\label{AsCCmST}
p_\mu \Gamma^{C^*\!CC}(p,q,r) + 2 q_\nu \Gamma^{A^*\!\!AC}_{\mu\nu}(p,q,r) = - C_A\left[ p_\mu \mathcal{B}(p) + r_\mu \mathcal{B}(r)\right] + \Delta^{A^*\!CC}_\mu(p,q,r)\,.
\ee
The $\Lambda$-integration constants satisfy the LHS alone 
and correspond to the freedom to change the normalization of the bracketed pair in \eqref{renormParam}. The rest of the above equation can be viewed as defining the longitudinal part 
$
\Gamma^{A^*\!\!AC}_{\mu\alpha}(p,q,r) P^L_{\alpha\nu}(q)
$
of $A^*\!AC$ vertex. Similarly, other mST relations define either longitudinal or transverse parts, via $Q_0 A_\nu$ or $Q^-_0 A^*_\mu$ respectively.

Inspecting fig. \ref{fig:Gamma3}, we see that there is only one diagram that contributes to $C^*C^2$ piece in \eqref{renormParam}. This is therefore the easiest way to extract $\gamma_g$ which in turn will give us the one-loop $\beta$-function. We find:
\be 
\label{gggc}
-\left(\gamma_g +\frac{\gamma_C}2\right)= \dot{\Gamma}^{C^*\!CC}(0,0,0) = \frac{C_A}2 \,\frac{\partial}{\partial t}\int_q \propH^2(q) q_\mu q_\nu \propH_{\mu\nu}(q) = \frac{\xi C_A}{(4\pi)^2}\int^\infty_0\!\!\!\!\!\! du\, \frac{\partial}{\partial u} \bar{K}^3(u) =  \frac{\xi C_A}{(4\pi)^2}\,.
\ee
To extract the $\beta$-function, we absorb the $Z^{-1}_g$ in \eqref{renormParam} into the coupling 
\be 
g(\Lambda) \, {\equiv g_{(r)} } = Z^{-1}_g g\,,
\ee
which thus runs. To one loop, the $\beta$-function is then
\be 
\label{beta}
\beta(g) = \Lambda \partial_\Lambda g(\Lambda) = - \dot{g}(\Lambda) = \gamma_g\, g^3(\Lambda)\,,
\ee
where from \eqref{gggc}, \eqref{gagc} and \eqref{ga} we recover the famous result, {here as a flow in $\Lambda$:}\footnote{The $\beta$-function under flow in $\Lambda$, specialised to SU$(2)$ and Feynman-gauge, was computed in ref. \cite{Bonini:1996bk}.}
\be 
\label{famous}
\gamma_g = \left(\gamma_g+\half \gamma_C\right) +\half \left(\gamma_A-\gamma_C\right) - \half\gamma_A = -\frac{11}{3}\frac{C_A}{(4\pi)^2}\,.
\ee

There are two diagrams that contribute to the $A^*\!AC$ vertex in \eqref{renormParam}, but otherwise the computation offers almost as straightforward a route to the $\beta$-function:
\beal
-\left(\gamma_g +\frac{\gamma_C}2\right)\delta_{\mu\nu} &= \dot{\Gamma}^{A^*\!\!AC}_{\mu\nu}(0,0,0)\nn\\ 
&=
\frac{C_A}2 \,\frac{\partial}{\partial t}\int_q \propH \left[ \propH q_\rho q_\nu \propH_{\mu\rho}
+q_\rho q_\sigma\propH_{\rho\sigma}\propH_{\mu\nu}+q_\rho q_\sigma\propH_{\mu\rho}\propH_{\nu\sigma}-2q_\nu q_\sigma\propH_{\mu\rho}\propH_{\rho\sigma}\right]\nn\\
&= \delta_{\mu\nu} \left(\xi+\xi(\xi+3)+\xi^2-2\xi^2\right)\frac{C_A}{4(4\pi)^2} \int^\infty_0\!\!\!\!\!\! du\, \frac{\partial}{\partial u} \bar{K}^3(u) =  \frac{\xi C_A}{(4\pi)^2} \delta_{\mu\nu}\,,
\label{gggcp}
\eeal
where the propagators are all evaluated at $q$. The end result agrees with \eqref{gggc} and thus again we get the famous one-loop $\beta$ function coefficient \eqref{famous}.  Verification of the wavefunction renormalization dependence of the other vertices in \eqref{renormParam} proceeds in a similar if somewhat longer fashion. {We note that through such relations, \cf \eqref{ga}, \eqref{gagc}, \eqref{gggc} and \eqref{gggcp}, we can confirm that the Slavnov-Taylor identities \eqref{SlavnovTaylor} are indeed satisfied.}

We finish by confirming explicitly that the body of the momentum integrals do automatically satisfy \eqref{AsCCmST}.  The $A^*C$ vertex in \eqref{G2q}, \cf \eqref{cal_B3}, can be written as
\be 
\label{AsC}
{-}C_{A} \propH_{\mu\nu} A^{*}_{\nu}\pa_{\mu}\propH C\,.
\ee
As we have seen, care is needed in defining the integration constant $\mathcal{B}_0\partial_\nu$ in this vertex, but here we will be interested in putting this to one side and demonstrating that the bulk of the integral already satisfies \eqref{AsCCmST}. For this we can proceed more formally. Using the notation in app. \ref{app:vertices},
the $C^*CC$ and $A^*AC$ vertices in \eqref{Gamma3} can similarly be written respectively as:
\be 
\label{threepoints}
\frac{i}{4} 
\pa_{\nu}\propH C^{*}\{\pa_{\mu} \propH C \propH_{\mu\nu},C\}\,,\qquad \frac{i}{2}C_{A} \propH_{\mu\nu} A_{\nu}^{*} [\pa_{\rho} 
\propH A_{\rho} \pa_{\mu}\propH,C]
+\frac{i}{2}C_{A} \pa_{\mu} \propH A_{\nu}^{*}[
{\cal T}_{\nu\mu}(A),C]\,.
\ee
The correction vertex $\Delta^{A^*\!CC}_\mu$ in \eqref{correction} and \eqref{AsCCmST}, takes the form
\be 
\label{correctionAsCC}
{+}\frac{i}{2}C_{A} K\Bigl( C\{\propH_{\mu\rho} A_{\rho}^{*} \pa_{\mu} \propH, C\} 
{+} C\{\propH_{\mu\rho} C \pa_{\rho} \propH,
A_{\mu}^{*}\}
{-} C\{\pa_{\mu}\propH C \propH_{\mu\nu},
A_{\nu}^{*}\}\Bigr)\,.
\ee
Using \eqref{AsC} and \eqref{sol1}, the antibracket part of \eqref{AsCCmST} can be written as
\be 
\label{antibracketAsCC}
{-}\frac{i}{2} C_{A} C\{\propH_{\mu\nu} A_{\mu}^{*} 
\pa_{\nu} \propH, C\}\ {+i} C_{A} C\{ \pa_{\nu} C 
\propH_{\mu\nu} \propH, A_{\mu}^{*} \}\,.
\ee
Finally operating on \eqref{threepoints}, the LHS of \eqref{AsCCmST}, namely $s_0\, \Gamma_{3,\text{q}} = (Q_0+Q^-_0) \Gamma_{3,\text{q}}$ in \eqref{mST3}, can be written as:
\be 
\label{sGammaAsCC}
-\frac{i}{4} C_{A}
\pa_{\nu}\propH \pa \cdot A^{*} 
\{\pa_{\mu} \propH C \propH_{\mu\nu},C\}
- \frac{i}{2}C_{A} \propH_{\mu\nu} A_{\nu}^{*} \{\pa_{\rho} 
\propH \pa_{\rho} C \pa_{\mu}\propH,C\}
-\frac{i}{2}C_{A} \pa_{\nu} \propH A_{\mu}^{*}\{
{\cal T}_{\mu\nu}(\pa C),C\}\,.
\ee
Using integration by parts, we find
\be 
{\cal T}_{\mu\nu}(\pa C) = (1-K)C\propH_{\mu\nu} - \propH_{\mu\nu}C(1-K) + \pa_{\mu}\pa_{\sigma}\propH
C \propH_{\sigma\nu} - \propH_{\mu\sigma} C \pa_{\sigma}\pa_{\nu}\propH \,,
\ee
and thus we find \eqref{sGammaAsCC} becomes:
\be 
\frac{i}{2}C_{A}(1-K) \Bigl(
C\{\pa_{\mu}\propH C \propH_{\mu\nu},A_{\nu}^{*}\}
- C\{\propH_{\mu\nu} C \pa_{\nu}\propH  ,A_{\mu}^{*}\}
- C\{\propH_{\mu\nu} A_{\mu}^{*}\pa_{\nu}\propH ,C\}
\Bigr)\,,
\ee
where use has been made of properties of the trace. Finally, substituting this, \eqref{correctionAsCC} and \eqref{antibracketAsCC} into \eqref{AsCCmST}, we find that this mST identity is indeed satisfied by the body of the momentum integrals. We remark that we have also confirmed the more complicated $AAC$ mST identity this way,  but do not report the details.

\section{Summary and Conclusions}
\label{sec:conclusions}

We reiterate the main points. The Wilsonian RG flow equation for the effective action $S$, can be combined with the QME \eqref{QMF} in such a way that they are mutually compatible \cite{Becchi:1996an,Igarashi:1999rm,Igarashi:2000vf,Igarashi:2001mf,Higashi:2007ax,Igarashi:2007fw,Sonoda:2007av,Sonoda:2007dj,Igarashi:2009tj,Frob:2015uqy}. A particularly natural formulation is given by the flow equation \eqref{flow} (which takes the same form as Polchinski's \cite{Polchinski:1983gv,Morris:1993}), with the QME regularised as in \eqref{QMEbitsReg} and the free action incorporating the free BRST transformations as in \eqref{S0} \cite{\morris}.
We have shown in general, and in detail for the example of perturbative Yang-Mills theory {with general gauge group and} in general gauge,  how for the continuum effective action solution, BRST invariance is then not broken by the presence of an effective ultraviolet cutoff but remains powerfully present in this structure, despite the fact that it demands quantum corrections that na\"\i vely break the gauge invariance, such as a mass term for the non-Abelian gauge field. 

In particular we have demonstrated that, combined with the derivative expansion property, BRST cohomological methods retain their potency. 
Key to this is the fact that the regularisation \eqref{QMEbitsReg} preserves the algebraic identities satisfied by the QME and its components, the antibracket and measure operator, and in such a way that the charges are then well defined when acting on arbitrary local functionals. Thus in particular the full quantum BRST charge $\hat{s}=Q+Q^--\Delta^--\Delta^=$, \cf \eqref{fullBRS} and \eqref{fullBRSsplit}, is well defined acting on arbitrary local functionals, and nilpotent off shell. 

The reason why terms which na\"ively should break the gauge invariance, are nevertheless allowed, in fact demanded, is precisely because the action of the Batalin-Vilkovisky measure operator is now non-trivial and well defined (in contrast to the usual treatment \cite{Gomis:1994he,Henneaux:1992ig}). Thus for example the one-loop two-point vertex $S_{2,\text{q}}$ satisfies 
\be
Q_0 \,S_{2,\text{q}} = (\Delta^-+\Delta^=)\, S_{2,\text{cl}}\,,
\ee 
as follows from \eqref{S2qRelations} and $Q_0^-S_{2,\text{q}}=0$, \cf \eqref{S2q} and \eqref{Q0-}. This
means that as well as any contribution that does satisfy linearised gauge invariance \eqref{Q0}, a part that is not $Q_0$-closed is required, which descends via the measure operator from the classical solution $S_\text{cl}$.

In practice it is the free full classical BRST charge $s_0 = Q_0 + Q^-_0$ and its non-trivial cohomology in the space of local functionals (on which it closes) that is key to developing the perturbative solution. Given $K$, such a solution is then unique up to the choice of renormalization conditions, {and thus is found without introducing \textit{a priori} either a classical action or a bare action.}
Indeed, thanks to the derivative expansion property, the $\Lambda$-integration constants we get from solving the flow equation for a finite effective action, are guaranteed by dimensions to be local functionals. Under very few assumptions, we can then recover crucial steps in {classical BRST cohomology \cite{Fisch:1989rp,Henneaux:1990rx,Barnich:1993vg,Barnich:1994db,Henneaux:1992ig,Henneaux:1997bm,Gomez:2015bsa} and in} the proof of renormalizability of gauge theories \cite{ZinnJustin:1974mc,ZinnJustin:1975wb,ZinnJustin:2002ru}. {This led} to the four central conclusions \ref{point1} -- \ref{point4} in sec. \ref{sec:five}, {which} state that the classical solution is {determined} by the non-trivial $s_0$-cohomology, with additional constraints arising from dimensional analysis. In particular \ref{point3} shows that if the CME is satisfied by such a solution to $O(g^3)$, then the CME is automatically satisfied at higher orders, while \ref{point4} concludes that the body of momentum integrals in the quantum corrections must already satisfy the QME. {As fully developed} in sec. \ref{sec:general}, we also listed the central conclusion \ref{point5}, namely that the RG flow under change of cutoff $\Lambda$ then generates a canonical transformation \eqref{canonicalS}--\eqref{wavefunctionGIS} which {automatically} solves the standard Slavnov-Taylor identities \eqref{SlavnovTaylor}
for the wavefunction renormalization constants (despite the presence of the cutoff). The remainder of the paper was devoted to developing and verifying these in detail for Yang-Mills theory {with general gauge group and} in general gauge.
And as we saw in sec. \ref{sec:oneloopsols} one then retrieves both the standard anomalous dimensions and the standard one-loop $\beta$-function (\ref{beta},\ref{famous}) {even though these are expressed} in terms of this cutoff.

Using the Legendre transformation identity \eqref{Legendre}
results in an equivalent one-particle-irreducible (1PI) description in terms of the simultaneous solution of the modified Slavnov-Taylor (mST) identities \eqref{mST} \cite{Ellwanger:1994iz}
and the flow \eqref{GammaFlow} of the 1PI  effective action $\Gamma$, the Legendre effective action {with} infrared cutoff $\bar{K}=1\cu-K$ {(\aka effective average action)} \cite{Nicoll1977,Morris:1993,Wetterich:1992}. These are written in terms of classical (anti)fields, whose antibracket \eqref{QMEbitsPhi} is the original one 
without the UV regularisation \cite{Batalin:1981jr,Batalin:1984jr,Batalin:1984ss}. Despite this, from \eqref{mST1} the free full BRST charge $\hat{s}_0$  operates in the same way, with the measure operator \eqref{DeltasG} still UV regularised by $K$. However
the presence of the quantum correction terms in the mST \eqref{mST} means that for $\Lambda\cu>0$, the analogous interacting full BRST charge is not nilpotent. In fact, as noted in sec. \ref{sec:discussion},  the classical (anti)fields and their antibracket are not canonically related to the quantum fields and respectively \eqref{QMEbitsReg} (although this implies that \eqref{mST} must however be subject to the non-linear symmetry relations, \cf app. \ref{app:mSTtrans}).

Nevertheless, the Legendre transformation identity guarantees that the perturbative development of the solution for $\Gamma$ mirrors that for $S$, as we verified explicitly in secs. \ref{sec:firstG} -- \ref{sec:oneloopG}. Indeed the freedom in the solution is again expressed through the $\Lambda$-integration constants which are guaranteed by dimensions to be local functionals. Then the key r\^ole is again played by the non-trivial cohomology of $s_0 = Q_0 + Q^-_0$ in the space of local functionals, where the free charges \eqref{BRSfree} generate the same algebra (\ref{Q0},\ref{Q0-}) as for the quantum fields. This means that the same conclusions \ref{point1} -- \ref{point5} hold for this 1PI formalism. In fact, since the antibracket \eqref{QMEbitsPhi} is not regularised, the classical $\Lambda$-integration constants are much more straightforward to determine, and the resulting classical solution \eqref{Gammacl} for $\Gamma_\text{cl}$ is a closed expression (unlike for $S_\text{cl}$) which coincides {with} the standard form of the classical action and its BRST transformations, {although it is derived rather than taken  \textit{a priori} as input.}
For the same reason the canonical transformation \eqref{canonicalK}--\eqref{wavefunctionGI} that provides the wavefunction renormalization factors parametrising the remaining freedom \eqref{renormParam} in the choice of $\Lambda$-integration constants, is much easier to determine in this formalism. Finally, and crucially, the $\Lambda\to0$ limit \eqref{limitisZJ} recovers the standard Legendre effective action satisfying the  Zinn-Justin equation \cite{ZinnJustin:1974mc,ZinnJustin:1975wb,ZinnJustin:2002ru,Ellwanger:1994iz}. As we confirmed in sec. \ref{sec:oneloopsols}, we thus recover the physical 1PI amplitudes, \ie the same ones one can compute by other methods modulo changes induced by differing renormalization scheme.

{In the early stage of the development of the subject, Yang-Mills theory
is studied with the exact RG method with antifields introduced as
sources for the BRST transformation \cite{Becchi:1996an}.  The
perturbative compatibility of the flow equation and QME is discussed in
\cite{Igarashi:2009tj} by studying the asymptotic behaviours of the
Wilsonian action and the QME.  In the present paper, we have derived an
explicit simultaneous solution to the flow equation and QME for a finite
cutoff at the one loop level in perturbation.}

While there is already a large literature on finding exact and approximate simultaneous solutions of the Legendre flow equation and mST for non-Abelian gauge theories, see \eg {\cite{Becchi:1996an,Igarashi:1999rm,Igarashi:2000vf,Igarashi:2001mf,Higashi:2007ax,Igarashi:2007fw,Sonoda:2007av,Sonoda:2007dj,Igarashi:2009tj,Frob:2015uqy,Ellwanger:1994iz,Bonini:1994kp,Bonini:1993sj,Reuter:1993kw,Ellwanger:1995qf,DAttanasio:1996jd,Bonini:1996bk,Litim1998,Freire:2000bq,Pawlowski:2005xe,Gies:2006wv,Fischer:2008uz,Mitter:2014wpa,Cyrol:2016tym,Cyrol:2017ewj},}  typically these approaches either lack control or become rapidly highly involved, depending especially on the extent to which the extra constraint from the mST is respected. We have shown how, through the derivative expansion property, $s_0$-cohomology (in the space of local functionals) facilitated by working in minimal gauge invariant basis (\ref{gibasisDeriv},\ref{GIbasisDeriv})  and exploiting antighost grading and descent equations \eqref{descendents}, both ensures the symmetry and allows elegant stream-lined derivations of renormalized perturbative solutions directly in the continuum, that simultaneously solve the flow equations and the mST. Moreover we have seen that the formulation in terms of the Wilsonian effective action $S$ is equivalent, and in this latter formulation the quantum BRST symmetry even at the interacting level is exact and well defined in the presence of the cutoff. 

On the other hand, the focus of part of the literature is in developing effective non-perturbative approximations \cite{Reuter:1993kw,Ellwanger:1995qf,Litim1998,Freire:2000bq,Pawlowski:2005xe,Gies:2006wv,Fischer:2008uz,Mitter:2014wpa,Igarashi:2016gcf,Cyrol:2016tym,Cyrol:2017ewj}.
In this paper, we have only applied these results explicitly to the development of exact perturbative solutions. It would be most interesting to investigate whether the structure allows a general framework for non-perturbative truncations  that continue to yield compatible flow equations and mST/QME identities and thus to allow approximations that yield non-perturbative simultaneous solutions. Even for a simple shift form of modified {Ward-Takahashi} identity, compatibility can be difficult to achieve in non-perturbative approximations \cite{Labus:2016lkh}, however we take heart from the considerable freedom in derivative expansion solutions of the QME alone (see the discussion at the beginning of sec. \ref{sec:secondWC}) and in the fact, noted below \eqref{AsCCmST}, that the mST may be viewed as a means to eliminate longitudinal and transverse parts of certain vertices\cite{Igarashi:2016gcf}.

We saw in sec. \ref{sec:compatibility} how the simultaneous solution of the flow equation and QME holds non-perturbatively in $\hbar$, defining the  continuum limit in terms of expansion over the eigenoperators and the quantum $\hat{s}_0$-cohomology within this space. This framework is important to progress to higher orders the  quantisation of gravity proposed by one of us \cite{Morris:2018mhd,Kellett:2018loq,Morris:2018upm,\morris}, which realises quantum gravity as a genuine continuum quantum field theory using interactions that are perturbative in Newton's constant but  non-perturbative in $\hbar$.
The equivalence under Legendre transformation will allow the 1PI formalism  also to be developed in this new quantisation scheme, so that the mST identities become satisfied at finite $\Lambda$, and the physical amplitudes are recovered in the $\Lambda\to0$ limit. 

\bigskip\bigskip

\section*{Acknowledgments}
{Y.I. and K.I. are supported by JSPS Grant-in-Aid for Scientific Research (C) JP19K03822.} 
T.R.M. acknowledges support during this research from a JSPS Bridge Fellowship BR150303, from the Leverhulme Trust and the Royal Society as a Royal Society Leverhulme Trust Senior Research Fellow, and from STFC through Consolidated Grants ST/L000296/1 and ST/P000711/1.


\appendix

\section{Further details and remarks}

\toclesslab\subsection{Relation to earlier formulations of QME and Exact RG}{app:shifted}

The basis introduced in ref. \cite{Morris:2018axr}  is related to the fields $\phi^A |_{unshifted}$ used in ref. \cite{Igarashi:2009tj} by a change of variables to the shifted fields $\phi^A$ discovered in refs. \cite{Higashi:2007ax,Igarashi:2007fw}. These latter fields appear naturally in the solution of the flow equation. The change of variables is in fact a (finite) quantum canonical transformation \cite{Morris:2018axr} (thus leaving the QMF invariant):
\be 
\label{me}
\phi^A  =  \phi^A |_{unshifted} + \propH^{AB} R^C_{\ B}\, \phi^*_C\,.
\ee
This basis leads to the advantage mentioned in the Introduction, namely that it provides us with a free quantum BRST algebra that closes on local functionals (since $\Delta$ in \eqref{QMEbitsReg} is well defined when acting on arbitrary local operators, while the free BRST transformations \eqref{Q0gen} are unmodified).
In fact the change of variables to shifted variables is forced on us at first order. See sec. 2.5 of \cite{Morris:2018axr} for more details.

\toclesslab\subsection{Translating between left and right acting  differentials}{app:leftright}

Right-acting free BRST and free Kozsul-Tate differentials are encoded into the free action \eqref{S0} by its last term. This should be compared to the case where they are left acting \cite{Morris:2018axr} \eg
\be 
Q\phi^A = (S,\phi^A)\,,
\ee
where we still require \eqref{Q0} and thus, thanks to the antibracket \eqref{QMEbitsReg}, we must have \cite{Morris:2018axr}
\be 
S_0 = \half \phi^A K^{-1} \prop^{-1}_{AB}\,\phi^B -R_{\ B}^A\phi^B K^{-1}\phi^*_A \,.
\ee
Comparing this with \eqref{S0}, we see that the translation corresponds simply to the replacement
\be 
R_{\ B}^A |_{left} = (-)^A R_{\ B}^A |_{right}\,.
\ee
For example, the relation \eqref{linearisedIdentity} then follows from translating eqn (A.9) in ref. \cite{Morris:2018axr}.

\toclesslab\subsection{Third order in coupling at classical level: notation}{app:notation}

As an example, suppose that parts of the action have the following form
\beal
\mathcal{S}_1 &=  [B,C]A \equiv i f^{abc} \int_x\! B^b(x)\, C^c(x)\, A^a(x)\,,\nn\\
\mathcal{S}_2 &=  D[E,F]\,,\nn\\
\mathcal{S}_3 &= 
G[H,J]\,, 
\eeal
where $A=A^a T^a$ \etc are fields in the adjoint representation which for simplicity we will assume to be Grassmann even. We use notation \eqref{notation}, and $f^{abc}$ are the group structure constants. Then at classical level, third order in the coupling, we encounter terms of the following form, where the notation is explained in full on the second line:
\besp
\left(\mathcal{S}_1\!\delr{A}\right)\propH \left(\dell{D}\mathcal{S}_2\!\delr{F}\right)\propH' \left(\dell{G} \mathcal{S}_4\right) = [B,C]\! \left[\, \propH\, E\, \propH'\,, [H,J] \right]\,,\\
\equiv i^3 \int_{x,y,z}\!\!\!\!\!\!f^{abc} B^b(x) C^c(x) \propH(x-y) f^{aef} E^e(y) \propH'(y-z) f^{fhj} H^h(z) J^j(z)\,.
\eesp

\toclesslab\subsection{Non-linear invariance of the mST}{app:mSTtrans}

The invariance of the QMF \eqref{QMF} under \eqref{gfgi} induces a transformation on the classical (anti)fields $\{\Phi,\Phi^*\}$ that must leave \eqref{mST} invariant since it is also $\Sigma$. Let us write \eqref{gfgi} more generally in the form
\be 
\label{bilinearGF}
\phi^*_A   = \check{\phi}^*_A  +\Psi_{AB}\phi^B\,.
\ee
This is a finite quantum canonical transformation (see app. A of \cite{Morris:2018axr}) generated by some generic bilinear `gauge fermion' $\Psi = \half \phi^A\Psi_{AB}\phi^B$ ($\Psi_{AB}$ being field independent, with $\epsilon_A =\epsilon_B+1$ and $\Psi_{AB} = \Psi_{BA}$). Using the Legendre transformation identity \eqref{LegendreIdentities}, \eqref{bilinearGF} can be recast as
\be 
\label{GIGFtrue}
\check{\Phi}^*_A  = \Phi^*_A - \Psi_{AB}\Phi^B - \Psi_{AB}\propH^{BC} \frac{\partial_l}{\partial\Phi^C}\Gamma_I[\Phi,\Phi^*]\,.
\ee
 Although this is a symmetry of \eqref{mST} which, by choosing $\Psi$ to be \eqref{Psi}, would take us to a gauge invariant basis $\check{\Psi}^*$, to apply it we would need to solve \eqref{GIGFtrue} for $\Phi^*$, yielding an infinite series in the form of a tree expansion.
 
\toclesslab\subsection{Equality of wavefunction renormalizations in the two effective actions}{app:Zs}

Expanding 
\be 
\label{sigma}
\Gamma_I[\Phi,\Phi^*] = \half \Phi^A\sigma_{AB}\Phi^B +\cdots\,,
\ee
where the ellipses stand for $\Phi^*$ terms and higher point vertices, 
one gets, either by solving \eqref{halfway} for the $S_I$ two-point vertex, or from the tree expansion, \cf \eqref{treeExpansion}, that 
\be 
S_I[\phi,\phi^*] = \half\, \phi^A\left( \sigma \left[1+\propH\sigma\right]^{-1}\right)_{AB}\!\phi^B +\cdots\,,
\ee
(Here one must recall \eqref{GIbasisDeriv} or work directly in gauge fixed basis.) Adding to this the free part in \eqref{S0} gives the modified kinetic term
\be 
S[\phi,\phi^*] = \half\,\phi^A K^{-1} \tilde{\prop}^{-1}_{AB} \phi^B +\cdots\,,
\ee
where 
\be 
\tilde{\prop}^{-1} = (\prop^{-1}+\sigma)[1+\propH\sigma]^{-1}\,.
\ee
On adding \eqref{sigma} to $\Gamma_0$, \cf \eqref{Gamma}, we set
\be 
\sigma = (Z^{-1} - 1) \prop^{-1} +\sigma'\,,
\ee
with $\sigma'(p)$ being $O(p^4)$, so as
to pull out the part parametrised by the field wavefunction renormalization constants \eqref{wavefunctionGI} (or rather as parametrised in \eqref{standard}). We thus have
\be 
\tilde{\prop}^{-1} = (Z^{-1}\prop^{-1}+\sigma')\left[1+(Z^{-1}-1)\bar{K}+\propH\sigma'\,\right]^{-1} = Z^{-1}\prop^{-1} + O(p^4)\,,
\ee
where the last equality follows since $\bar{K}$ is $O(p^2)$. (In fact, recalling \eqref{Kprime}, it is even $O(p^4)$.) This establishes that the wavefunction renormalization constants for $\Gamma$ work in $S$ and take the same value. A similar analysis can be carried through to verify this property also for the $\phi^*$ terms.

\toclesslab\subsection{Defining the conditionally convergent parts}{app:Care}

The demonstration that the $K$-independent part \eqref{vanish} vanishes
 and similar manipulations, need care because \eqref{ALplus} is only conditionally convergent, the convergence arising from cancellation of divergent terms. Consider for example the following integral that arises from extracting the $K$ independent part:
\be 
\label{cubic}
4 \int_q \frac{p\cu\cdot q}{p^2q^2}\,.
\ee
This integral vanishes by Lorentz invariance. However since it appeared multiplied by $\bar{K}(q)\bar{K}(p\cu+q)$ in \eqref{ALplus}, we could equally have written it as:
\be 
\label{quadratic}
2 \int_q \left( \frac{p\cu\cdot q}{p^2q^2} - \frac{p\cu\cdot (q+p)}{p^2(q\cu+p)^2} \right)  = 
2\int_q \frac{p\cu\cdot q (2p\cu\cdot q+p^2)-q^2p^2 }{p^2q^2(q\cu+p)^2}\,.
\ee
This latter integral now has a non-vanishing quadratic divergence (which one can readily extract by taking the  $p\to0$ limit). The problem is that \eqref{cubic} is superficially cubically divergent. The non-vanishing contribution \eqref{quadratic} is in fact a surface term, which cannot be dropped since it is not only non-vanishing but quadratically divergent. 

However the final version, which coincides with \eqref{F} is well defined. If the intermediate expressions are treated in the same way, the result is also unambiguous. Thus for example if we start with the expression \eqref{ALplus} but with the ${p\cu\cdot q}/{p^2q^2}$ replaced by the integrand in \eqref{quadratic}, then we would now need a divergent integration constant chosen to cancel the quadratic divergence. However when we finish, we would again get \eqref{F}. Alternatively we could also supply a UV regularisation, as we have to do anyway for $\mathcal{B}$ for example in sec. \ref{sec:oneloopsecondorder}, after which the divergent part, and thus corresponding integration constant, is unambiguous.

Likewise, the third term in brackets in \eqref{Bintegral} requires a little care in taking the $p\to0$ limit. Using $q\mapsto-q-p$ symmetry we have that
\be 
\frac{q\cu\cdot p}{p^2q^4} \equiv \frac12 \left( \frac{q\cu\cdot p}{p^2q^4} - \frac{q\cu\cdot p + p^2}{p^2(q\cu+p)^4}\right)\,.
\ee
Combining over a common denominator, and expanding the numerator, we can neglect all terms that evidently vanish as $p\to0$. We are then left with
\be
\frac{(q\cu\cdot p)^2}{p^2q^6} -\frac1{4\,q^4}
\ee
which clearly provides a vanishing contribution to \eqref{Batz}.

\toclesslab\subsection{Vertices for one-loop contributions}{app:vertices}

Using the alternative to \eqref{GIbasisDeriv}, we write \eqref{sol1} as:
\be 
\Gamma_{1} = f^{abc} \int \Bigl[\pa_{\mu}A^{a}_{\nu}A^{b}_{\mu}A^{c}_{\nu} + 
\frac{1}{2} C^{*a}C^{b}C^{c} + \bigl(A^{*}_{\mu}- i \pa_{\mu}{\bar C}\bigr)^{a}
A_{\mu}^{b}C^{c}\Bigr] \,.
\ee
Defining $\tau_{AB}^{C} = \dell{A} \Gamma_{1} \delr{B}$, and writing
\bea
&&\bigl(\tau_{\mu c}^{A^{*}}\bigr)^{ab}(x,y) \equiv \frac{\pa}{\pa A_{\mu}^{a}(x)}
\frac{\pa^{r}}{\pa C^{b}(y)}\Gamma_{1} = -f^{acb}A_{\mu}^{*c}(x)\delta(x-y) 
= \bigl(\tau_{c \mu }^{A^{*}}\bigr)^{ab}(x,y)\,,
\nn\\
&& 
\bigl(\tau_{cc}^{C^{*}}\bigr)^{ab}(x,y) = \frac{\pa^{l}}{\pa C^{a}(x)}
\frac{\pa^{r}}{\pa C^{b}(y)}\Gamma_{1} = -f^{acb}C^{*c}(x)\delta(x-y)\,,
\label{vertex}\\
&& 
\bigl(\tau_{\mu\nu}^{A}\bigr)^{ab}(x,y) = \frac{\pa}{\pa A_{\mu}^{a}(x)}
\frac{\pa}{\pa A_{\nu}^{b}(y)}\Gamma_{1} = f^{acb} \tau_{\mu\nu}^{A^{c}}(x,y)~,
\qquad
 \bigl(\tau_{{\bar c}c}^{A}\bigr)^{ab}(x,y) 
= \frac{\pa^{l}}{\pa {\bar C}^{a}(x)}
\frac{\pa^{r}}{\pa C^{b}(y)}\Gamma_{1} \,,
\nn
\end{eqnarray}
for derivative couplings connected to propagators on either side, we have
\beal
 \prop\bigl(\tau_{{\bar c}c}^{A}\bigr)^{ab}\prop &= 
{+i} f^{acb} \pa_{\mu}\prop A^{c}_{\mu} \prop\,,
\quad
\prop\bigl(\tau_{{c\bar c}}^{A}\bigr)^{ab}\prop = -i f^{acb} \prop A^{c}_{\mu}\pa_{\mu} \prop\,,\nn\\
\prop\bigl(\tau_{{\bar c}\mu}^{C}\bigr)^{ab} \prop &= {-i} f^{acb} \pa_{\mu}\prop C^{c} \prop\,,
\quad
\prop\bigl(\tau_{{\mu\bar c}}^{C}\bigr)^{ab}\prop = i f^{acb} \prop
 C^{c}\pa_{\mu} \prop\,,
\label{dervertex2}
\eeal
with similar expressions for the $\Gamma^{(2)}_{1*}$ vertices. Defining ${\cal T}_{\rho\sigma}(A^c) = \prop_{\rho\mu}\bigl(\tau_{\mu\nu}^{A^{c}}\bigr)\prop_{\nu\sigma}$, we find
\be
\label{T}
{\cal T}_{\rho\sigma} (A^c)
= \Bigl[2\pa_{\nu}\prop_{\rho\mu}A^{c}_{\mu} + 2\prop_{\rho\mu} A^{c}_{\nu}
\pa_{\mu} - \prop_{\rho\mu}A^{c}_{\mu}\pa_{\nu} - \prop_{\mu\nu}\bigl(\pa \prop_{\rho\mu}\cdot A^{c}  
+ \prop_{\rho\mu}A^{c}\cdot \pa \bigr) - \xi \pa_{\rho}\prop A_{\nu}^{c}
\Bigr]\prop_{\nu\sigma}\,.
\ee

\toclesslab\subsection{Gauge field double propagator terms in the AA vertex}{app:XYZ}

Applying the projectors \eqref{Asplit} we have
\be 
\mathcal{A}^L = \mathcal{A}_{\mu\nu}P^L_{\mu\nu}\qquad\text{and} \qquad \mathcal{A}^T = \tfrac13\mathcal{A}_{\mu\nu} P^T_{\mu\nu} =\tfrac13\left(\mathcal{A}_{\mu\mu} -\mathcal{A}^L\right)\,.
\ee
The most involved part, the gauge field double propagator piece, of \eqref{A}:
\be
\label{Agaugedouble}
-\half\,  \propH_{\alpha\beta}(q)\propH_{\rho\sigma}(q\cu+p)\,\Theta_{\rho\alpha\mu}\Theta_{\sigma\beta\nu} \,,
\ee
gives for $I=L,T$ contribution
\be 
-\half\,\propH(q)\propH(q\cu+p) \left( X^I +\xi Y^I+ \xi^2 Z^I\right)\,,
\ee
where the expressions $X^I$, $Y^I$ and $Z^I$ are $q\mapsto-q\cu-p$ symmetric:
\beal
 X^{T} 
&= 3\Bigl[(p-q)^{2} - \frac{(p^{2}-q^{2})^{2}}{(p+q)^{2}}\Bigr]
+ 3\Bigl[(2p+q)^{2} - \frac{(2 p\cdot q +q^{2})^{2}}{q^{2}}\Bigr]
+ (p + 2q)^{2} \Bigl[2 + \frac{(p\cdot q +q^{2})^{2}}{q^{2}(p+q)^{2}}\Bigr]\,,
\nn\\
 Y^{T}
&= (p-q)^{2} + 2 (p + 2q)^{2} -3q^{2} + 3p\cdot q +  (2p+q)^{2}
-3 (p+q)^{2} -3(p^{2} +p\cdot q )\nn\\ &+ 2 \frac{(2 p\cdot q +q^{2})^{2}}{q^{2}} 
+2 \frac{(p^{2}-q^{2})^{2}}{(p+q)^{2}} 
+ 3 \Bigl[\frac{(p\cdot q +q^{2})^{2}}{q^{2}} + \frac{(p\cdot q)(p\cdot q +q^{2})}{q^{2}} \Bigr] \nn\\
&
+3 \Bigl[\frac{(p\cdot q +q^{2})^{2}}{(p+q)^{2}} 
- \frac{(p\cdot q +q^{2})(p^{2} + p\cdot q)}{(p+q)^{2}}\Bigr]
-2 (p + 2q)^{2} \frac{(p\cdot q +q^{2})^{2}}{q^{2}(p+q)^{2}} \,,
\nn\\
 Z^{T}
&= \frac{(p^{2}-q^{2})^{2}}{(p+q)^{2}} + \frac{(2 p\cdot q +q^{2})^{2}}{q^{2}} 
+ (p + 2q)^{2} \frac{(p\cdot q +q^{2})^{2}}{q^{2}(p+q)^{2}} 
\nn\\
&
- 3 \Bigl[\frac{(p\cdot q) (p\cdot q +q^{2})}{q^{2}} +(p\cdot q +q^{2})^{2}
\Bigl\{\frac{1}{q^{2}} + \frac{1}{(p+q)^{2}}\Bigr\} 
- \frac{(p\cdot q +q^{2})(p^{2} + p\cdot q)}{(p+q)^{2}}\Bigr]\,,
\nn\\
 X^{L} &= (p^{2}+2p\cdot q)^{2}\Bigl[ 2 
+ \frac{(p\cdot q+q^{2})^{2}}{(p+q)^{2}q^{2}}\Bigr] \,,
\label{XYZL}\\
Y^{L} 
&= (p+q)^{2}\Bigl\{p^{2}-
\frac{(p\cdot q)^{2}}{q^{2}}\Bigr\}+ q^{2}\Bigl\{p^{2}-
\frac{(p^{2}+p\cdot q)^{2}}{(p+q)^{2}} \Bigr\}\,,
\nn\\
Z^L &= 0\,.
\nn
\eeal


\bibliographystyle{hunsrt}
\bibliography{references} 

\end{document}